\documentclass[a4paper]{article}
\usepackage{amsmath,amssymb,amsthm,amscd}
 \def\lb{\label}

\def\bc{\begin{center}}
\def\ec{\end{center}} \DeclareMathOperator{\Ad}{Ad} \DeclareMathOperator{\ad}{ad}
\DeclareMathOperator{\Diff}{Diff}\DeclareMathOperator{\const}{const}
\DeclareMathOperator{\LDiff}{LDiff}
\DeclareMathOperator{\RDiff}{RDiff}
\DeclareMathOperator{\LRDiff}{LRDiff}
\DeclareMathOperator{\arctanh}{arctanh}
\DeclareMathOperator{\id}{id} 
 
 \DeclareMathOperator{\diag}{diag}
 \DeclareMathOperator{\diam}{diam}

 \newtheorem{theore}{Theorem}
\newtheorem{Def}{Definition}
\newtheorem{proposit}{Proposition}
\newtheorem{Rem}{Remark}
\newtheorem{Lem}{Lemma}
\newtheorem{Con}{Condition} \def\pd#1#2{\frac{\partial{#1}}{\partial{#2}}}
\def\pd1#1{\frac{\partial}{\partial#1}}
\def\d1#1{\frac{d}{d#1}}

\begin{document} \author{A.V. Shchepetilov\\ Department of Physics, Moscow State University, \\ 119992 Moscow, Russia,
email: alexey@quant.phys.msu.su}

\title{Two-body problem on two-point homogeneous spaces, invariant differential operators
and the mass center concept}
\date{}
\maketitle
\begin{abstract} We consider
the two-body problem with central interaction on two-point
homogeneous spaces from the point of view of the invariant
differential operators theory. The representation of the
two-particle Hamiltonian in terms of the radial differential
operator and invariant operators on the symmetry group is found.
The connection of different mass center definitions for these
spaces to the obtained expression for Hamiltonian operator is
studied.
\end{abstract}

\section{Introduction} \label{Introduction}
\markright{\ref{Introduction} Introduction}

The purpose of this paper is to provide a comprehensive treatment
of the quantum-mechanical two-body problem on Riemannian two-point
homogeneous spaces from the point of view of the theory of
invariant differential operators developed by S. Helgason
\cite{Hel1} and others (see \cite{Hel85}-\cite{Vinberg} and
references therein).

Let $M$ be a Riemannian manifolds with an action of an isometry
group $G$ on it. We assume that $G$-orbits in $M$ of a maximal
dimension $\ell$ are isomorphic to each other, their union is an
open dense submanifold $M'$, the set $M\backslash M'$ has zero
measure, and $M'=W\times\mathcal{O}$, where $\mathcal{O}$ is a
$G$-orbit of a maximal dimension, and $W$ is a submanifold of $M'$
transversal to all $G$-orbits of the dimension $\ell$. This
situation is a typical one \cite{Bred}, and we have the
isomorphism of measurable sets $(M,\mu)\cong
(W,\nu)\times(\mathcal{O},\mu_{G})$, where $\mu$ is the
$G$-invariant measure on the manifold $M$, generated by its
metric, $\nu$ is some measure on $W$, and $\mu_{G}$ is a
$G$-invariant measure on $\mathcal{O}$. It implies the
isomorphisms (see for example Theorem II.10 in \cite{SR}):
\begin{equation}\lb{isomor}
\mathcal{H}:=\mathcal{L}^{2}(M,d\mu)=\mathcal{L}^{2}(W\times\mathcal{O},d\nu\otimes
d\mu_{G})=\mathcal{L}^{2}(W,d\nu)\otimes\mathcal{L}^{2}(\mathcal{O},d\mu_{G}).
\end{equation} Under these assumptions, an invariant differential operator $D$ on
$M'$ admits an explicitly symmetric decomposition of the form:
\begin{equation}
\lb{MainExpansion} D=D_{T}+\sum_{(i)}D_{(i)}\circ
X_{i_{1}}^{+}\circ\dots\circ X_{i_{r}}^{+}\equiv
D_{T}+\sum_{(i)}D_{(i)}\circ\square_{(i)},
\end{equation}
where $X_{i}^{+}$ is a differential operator of the first order,
corresponding to the action of the one parametric subgroup $\exp
(tX_{i}),\; X_{i}\in\mathfrak{g}$, of the group $G$ on the space
$M'$; $D_{T}$ and $D_{(i)}$ are transversal operators with respect
to some manifold $W$ which is in transverse position with respect
to orbits of the group $G$ in $M'$; here $D_{T}$ is called the
transversal part of $D$ (see theorem 3.4, chapter II,
\cite{Hel85}). Operators $\square_{(i)}$ are invariant on orbits
of the group $G$ in space $M'$. Such operators can be naturally
expressed in terms of the Lie algebra $\mathfrak{g}$ of the group
$G$. In this paper we assume $M$ and $G$ to be connected. Such an
expression of invariant differential operators corresponds to a
general approach to invariant differential geometrical objects on
homogeneous spaces. These objects have the simplest form in the
basis of Killing vector fields \cite{Besse}, \cite{KoNo1}. For
invariant metrics on Lie groups this approach was developed in
\cite{Arn1} in the direction to infinite-dimensional groups. Note
that the representation (\ref{MainExpansion}) depends on a choice
of a transversal manifold $W$. Now let an operator $D=H$ be a
Hamiltonian of some quantum mechanical system on a Riemannian
homogeneous manifold $M$ acting in the space $\mathcal{H}$.

The group $G$ naturally acts on the space
$\mathcal{L}^{2}(\mathcal{O},d\mu_{G})$ by left-shifts, and if
$\mathcal{H}'$ is an invariant subspace of the latter space, then
the operator $H$ admits  a restriction on the space
$\mathcal{L}^{2}(W,d\nu)\otimes\mathcal{H}'$. For a compact group
$G$ we can expand the space
$\mathcal{L}^{2}(\mathcal{O},d\mu_{G})$ into the direct sum of
spaces of irreducible representations of the group $G$ and obtain
from (\ref{isomor})
$$
\mathcal{H}=\mathcal{L}^{2}(W,d\nu)\otimes\left(\oplus_{i}\mathcal{H}'_i\right)=
\oplus_{i}\left(\mathcal{L}^{2}(W,d\nu)\otimes\mathcal{H}'_i\right).
$$  In this case the Hamiltonian $H$ is expanded in the direct
sum of operators in spaces
$\mathcal{H}_{i}=\mathcal{L}^{2}(W,d\nu)\otimes\mathcal{H}'_{i}$.
Hence the symmetric quantum mechanical system is reduced to the
set of subsystems that are not $G$-symmetric. This method was
described in papers \cite{Landsman} -- \cite{TanIwai} without
mentioning the expansion (\ref{MainExpansion}). On the other hand,
it seems to be difficult to manage without the expansion
(\ref{MainExpansion}) while reducing the quantum mechanical
system, because representation theory for the group $G$ gives only
the formulae for the action of operators $\square_{(i)}$ in the
space $\mathcal{H}'_{i}$. Without (\ref{MainExpansion}), the
calculation of the action of $H$ in the space $\mathcal{H}_{i}$
requires cumbersome computations.

At the same time, the expansion (\ref{MainExpansion}) gives some
information on the complexity of reduced subsystems even in the
absence of the detailed information about irreducible
representations of the group $G$ in the space
$\mathcal{L}^{2}(\mathcal{O},d\mu_{G})$. For example, if all
operators $\square_{(i)}$ in (\ref{MainExpansion}) commute, they
have only common eigenfunctions, and the spectral problem for the
Hamiltonian $H$ reduces to a set of spectral problems for some
scalar differential operators on the manifold $W$.

Now let $H=H_{0}+U$ be a Hamiltonian of the system of two
particles on a Riemannian space $Q$. Here
$H_{0}=-\dfrac1{2m_{1}}\triangle_{1}
-\dfrac1{2m_{2}}\triangle_{2}$ is the free two-particle
Hamiltonian (everywhere we put $\hbar=1$), $m_{1}, m_{2}$ are
particle masses, $\triangle_{i},\,i=1,2$ is the Laplace-Beltrami
operator on the $i$-th factor of the configuration space
$M=Q\times Q$ for this system, $U$ is the interaction potential
depending only on a distance between particles, and $G$ is the
identity component of an isometry group of the space $Q$. The
group $G$ acts naturally on the space $Q\times Q$ as
$$ g:\; (q_{1},q_{2})\rightarrow(gq_{1},gq_{2}),\,g\in G,\,
(q_{1},q_{2}) \in Q\times Q.
$$
The dimension of a  manifold $W\subset M$ in this case is one, or
greater, since the group $G$ conserves a distance between two
points of the space $Q$. In other words, the codimension of
$G$-orbits in $M$ is one or greater. In this paper we consider the
case of two-point homogeneous Riemannian spaces $Q$ for which the
latter codimension is equal to one. On the spaces of constant
sectional curvature, which are the special case of two-point
homogeneous Riemannian spaces, this problem was considered in
\cite{Shchep}--\cite{ShchStep}.

This paper is organized as follows. In  section \ref{InvOperators}
we consider the theory of invariant differential operators
emphasizing the facts which will be used later for calculating the
two-point Hamiltonian. In section \ref{LaplaceRepere} we find the
formula for the Laplace-Beltrami operator in a basis of Killing
vector fields. This formula for the basis consisting of Killing
vector fields and a radial vector field is then generalized in
section \ref{hom}. The classification of two-point homogeneous
Riemannian spaces is given in section \ref{twopoint}. There is
also a construction of a special basis for the Lie algebra
$\mathfrak{g}$ of the group $G$. We use this construction and the
formula for the Laplace-Beltrami operator from
 section \ref{hom} to obtain the expression of the type
(\ref{MainExpansion}) for the two-particle Hamiltonian on
two-point compact homogeneous Riemannian spaces in section
\ref{generaltwobody}. Using the formal correspondence between
compact and noncompact two-point homogeneous spaces, in section
\ref{noncompact} we transform the latter expression into the form
valid for the noncompact case. The Hamiltonian two-particle
functions for two-point homogeneous spaces are considered in
section \ref{ClassicalSystems}. Different mass center concepts on
two-point homogeneous spaces are discussed in section
\ref{centermass}. We study the connection of the mass center
concepts to the obtained expressions for quantum and classical
Hamiltonians. \section{Invariant differential operators on
homogeneous spaces} \label{InvOperators}
\markright{\ref{InvOperators} Invariant differential operators}
Let $M$ be a Riemannian  $G$-homogeneous space, $\dim M=m$, $\dim
G=N, x_{0}\in M, K_{x_{0}}\subset G$ a stationary subgroup of a
point $x_{0}\in M$, and $\mathfrak{k}_{x_{0}}\subset
\mathfrak{g}\equiv T_{e}G$ the corresponding Lie algebras. Choose
a subspace $ \mathfrak{p}_{x_{0}}\subset\mathfrak{g}$ such that
$\mathfrak{g}=\mathfrak{p}_{x_{0}}\oplus\mathfrak{k}_{x_{0}}$(a
direct sum of linear spaces).

The stationary subgroup $K_{x_{0}}$ is compact, since it is also
the subgroup of the group $\mathbf{SO}(m)$. By the group averaging
on $K_{x_{0}}$ we can define a $\Ad_{K_{x_{0}}}$-invariant scalar
product on $ \mathfrak{g}$ and choose the subspace
$\mathfrak{p}_{x_{0}}$ orthogonal to $\mathfrak{k}_{x_{0}}$ with
respect to this product \cite{Hel85}, \cite{KoNo2}. In this case
we have $\Ad_{K_{x_{0}}}(\mathfrak{p}_{x_{0}})\subset
\mathfrak{p}_{x_{0}}$, i.e. the space $M$ is reductive.

Identify the space $M$ with the factor space of left conjugate
classes of the group $G$ with respect to the subgroup $K_{x_{0}}$.
Let $\pi
:G\to G/K_{x_{0}}$ be the natural projection. Denote by
$$
L_{q}:q_{1}\rightarrow qq_{1},\quad R_{q}: q_{1}\rightarrow
q_{1}q,\quad q,q_{1}\in G
$$
the left and the right shifts on the group $G$, and by
$$
\tau_{q}: x\rightarrow qx,\quad q\in G,\quad x\in M
$$
the action of an element $q\in G$ on $M$. Obviously, $\pi\circ
L_{q}=\tau_{q}\circ\pi,\quad q\in G $ ¨ $\pi \circ R_{q}=\pi,\quad
q\in K_{x_{0}}$. Let the left and the right shifts act on the
space $C^{\infty}(G)$ as:
$$
\hat L_{q}(f)(q_{1})=f(q^{-1}q_{1}),\qquad \hat
R_{q}(f)(q_{1})=f(q_{1}q^{-1}),\quad f\in C^{\infty}(G);
$$
The left shift acts on the space $C^{\infty}(M)$ as:
$$
\hat \tau_{q}(f)(x)=f(q^{-1}x),\quad f\in C^{\infty}(M).
$$
Then $\hat L_{q_{1}q_{2}}=\hat L_{q_{1}}\circ \hat L_{q_{2}},\quad
\hat R_{q_{1}q_{2}}=\hat R_{q_{2}}\circ \hat R_{q_{1}},\quad\hat
\tau_{q_{1}q_{2}}=\hat\tau_{q_{1}}\circ\hat\tau_{q_{2}},\,\hat
L_{q_{1}}\circ\hat R_{q_{2}}=\hat R_{q_{2}}\circ\hat L_{q_{1}},\,
q_{1},q_{2}\in G $.

Let $\Diff(G)$ and $\Diff(M)$ be algebras of differential
operators with smooth coefficients on $G$ and $M$, respectively.
Define the action of shifts on operators as:
\begin{eqnarray*}
\tilde L_{q}(\square)=\hat L_{q}\circ\square\circ \hat
L_{q^{-1}},\quad \tilde R_{q}(\square)=\hat R_{q}\circ\square\circ
\hat R_{q^{-1}},\quad \square\in\Diff(G),\\
\tilde\tau_{q}(\square)=\hat\tau_{q}\circ\square\circ\hat
\tau_{q^{-1}},\quad \square\in\Diff(M).
\end{eqnarray*}
Define the following subalgebras:
$$
\LDiff(G):=\left\{\square\in\Diff(G)|\,\tilde L_{q}(
\square)=\square,\, \forall q\in G\right\},
$$
$$
\LDiff(M):=\left\{\square\in\Diff(M)|\,\tilde \tau_{q}(
\square)=\square,\, \forall q\in G\right\},
$$
$$
\RDiff(G):=\left\{\square\in\Diff(G)|\,\tilde R_{q}(
\square)=\square,\,\forall q\in G\right\},
$$
$$
\LRDiff(G):=\left\{\square\in\LDiff(G)|\,\tilde R_{q}(
\square)=\square,\,\forall q\in G\right\},
$$
$$
\LDiff_{K}(G):=\left\{\square\in\LDiff(G)|\,\tilde R_{q}(
\square)=\square,\,\forall q\in K\right\},
$$
where $K$ is a subgroup of $G$. For any algebra $\mathcal{A}$
denote $\text{Z}\,\mathcal{A}$ the center of $\mathcal{A}$. Let
$S(V)$ be a symmetric algebra over a finite dimensional complex
space $V$, i.e. a free commutative algebra over the field
$\mathbb{C}$, generated by elements of any basis of $V$. The
adjoint action of the group $G$ on $\mathfrak{g}$ can be naturally
extended to the action of $G$ on the algebra $S(\mathfrak{g})$
according to the formula:
$$
\Ad_{q}:\,Y_{1}\cdot\ldots\cdot Y_{i}\rightarrow
\Ad_{q}(Y_{1})\cdot\ldots\cdot\Ad_{q}(Y_{i}),\,Y_{1},\dots,Y_{i}\in
\mathfrak{g}.
$$
Denote by $I(\mathfrak{g})$ the set of all $\Ad$-invariants in
$S(\mathfrak{g})$.

Let $e_{m+1},\dots,e_{N}$ be a basis in $\mathfrak{k}$, and
$e_{1},\dots,e_{N}$ a basis in $\mathfrak{g}$. There are
corresponding moving frames on the group $G$ consisting
respectively of the following left and right invariant vector
fields:
$$
X_{i}^{l}(q)=dL_{q}e_{i},\,\quad X_{i}^{r}(q)=dR_{q}e_{i},\,
i=1,\dots,N,\, q\in G.
$$
There are also the dual moving frames $X^{i}_{l}(q),
X^{i}_{r}(q)$. In general, we shall denote by $Y^{l}$ and $Y^{r}$
the left- and the right-invariant vector fields, corresponding to
an element $Y\in\mathfrak{g}$.

We can consider vector fields as differential operators of the
first order and any differential operator on the group $G$ can be
expressed as a polynomial in $X_{i}^{l}$ or in $X_{i}^{r},\,
i=1,\dots,N$ with nonconstant coefficients. Define a map
$$
\lambda :\,S( \mathfrak{g})\rightarrow\Diff(G)
$$
by the formula
\begin{equation*}
\left( \lambda(P)f\right)(q)=\left[P(
\partial_{1},\ldots,\partial_{N})f\left(q\exp(t_{1}e_{1}+\ldots+t_{N}e_{N})\right)\right]_{t=0},
\end{equation*}
where $P\in S( \mathfrak{g})$ on the left-hand side is a
polynomial in $e_{1},\ldots e_{N}$ and on the right-hand side the
substitution $e_{i}\rightarrow\partial_{i}:=\partial/\partial
t_{i},\,i=1,\dots,N$ was made. Here $
t=(t_{1},\ldots,t_{N})\in\mathbb{R}^{N},\, f\in C^{\infty}(G)$.
\begin{theore}[\cite{Hel85}]
\label{T1} The map $\lambda$ is the unique linear bijection
(generally not a homomorphism) of the algebra $S(\mathfrak{g})$
onto the algebra $\LDiff(G)$ such that
\begin{equation*}
\lambda((Y)^{i})=(Y^{l})^{i}=\underbrace{Y^{l}\circ\ldots\circ
Y^{l}}_{\mbox{$i$ times}},\, Y\in \mathfrak{g}.
\end{equation*}
\end{theore}
\begin{Rem}
\label{Rem1} The map $\lambda$ transforms the element
$Y_{1}\cdot\ldots\cdot Y_{p}\in S(\mathfrak{g})$ into the operator
$$
\frac{1}{p!}\sum\limits_{ \sigma\in \mathfrak{S}
_{p}}Y_{\sigma(1)}^{l}\circ\ldots\circ Y_{\sigma(p)}^{l}\in
\LDiff(G),
$$
where $\mathfrak{S}_{p}$ is the group consisting of all
permutations of $p$ elements. The map $\lambda$ is called
symmetrization. With its help the noncommutative algebra
$\LDiff(G)$ is described in terms of the free commutative algebra
with $N$ generators $e_{1},\ldots,e_{N}$.
\end{Rem} Let $\tilde{\Ad}_{q}\square:=\tilde L_{q}\circ\tilde
R_{q^{-1}}(\square),\,\square\in\Diff(G),\, q\in G$. It is clear,
that $\tilde{\Ad}_{q}$ is the automorphism of algebras
$\Diff(G),\,\LDiff(G),\,\RDiff(G)$. Obviously,
\begin{equation*}
\tilde{\Ad}_{q}\square=\tilde
R_{q^{-1}}(\square),\,\square\in\LDiff(G).
\end{equation*}
By direct calculations we have:
\begin{equation}
\label{Ad2} \left(\Ad_{q}Y\right)^{l}=\tilde{\Ad}_{q}Y^{l}=\tilde
R_{q^{-1}}Y^{l},\,\forall Y \in \mathfrak{g}.
\end{equation}
Define the operation
\begin{equation*}
\tilde{\ad}_{Y}\square=Y^{l}\circ\square-\square\circ
Y^{l},\,\forall Y \in \mathfrak{g}, \square\in\Diff(G).
\end{equation*}
Then
\begin{equation}
\label{Ad3} \left(\ad_{Y}X\right)^{l}=Y^{l}\circ X^{l}-X^{l}\circ
Y^{l}=\tilde{\ad}_{Y}X^{l},\, X,Y \in \mathfrak{g}.
\end{equation}
Evidently, the operation $\tilde{\ad}$ is a differentiation of
algebras $\Diff(G)$ ¨ $\LDiff(G)$. By direct calculations we
conclude that the operation
\begin{equation*}
\exp(\tilde{\ad}_{Y})D:=\sum\limits_{i=0}^{ \infty
}\frac{1}{i!}\tilde{\ad}^{i}_{Y}(D),\,D\in\Diff(G),
\end{equation*}
is the automorphism of algebras  $\Diff(G)$ and $\LDiff(G)$. It is
well known that the operations $\exp(\ad_{Y})$ and $\Ad_{\exp(Y)}$
coincide on $\mathfrak{g}$, so using (\ref{Ad2}) and (\ref{Ad3})
we see that the operations $\tilde{\Ad}_{\exp Y}$ and
$\exp(\tilde{\ad}_{Y})$ coincide also on operators
$X^{l}\in\LDiff(G),\,X\in \mathfrak{g}$. Since the operators
$\tilde{\Ad}_{\exp Y},\,\exp(\tilde{\ad}_{Y})$ are automorphisms
of the algebra $\LDiff(G)$ and the operators $X_{i}^{l}$ are the
generators of the algebra $\LDiff(G)$ according to Theorem
\ref{T1}, the equality
\begin{equation*}
\tilde{\Ad}_{\exp Y}=\exp(\tilde{\ad}_{Y}),
\end{equation*}
holds everywhere in $\LDiff(G)$.

Functions on space $M$ are in one to one correspondence with the
functions on group $G$ that are invariant under the right action
of the subgroup $K_{x_{0}}$. This correspondence is defined by the
formula $\zeta: f\to \tilde f:=f\circ\pi$, where $f$ is a function
on space $M$ and  $\tilde f$ the corresponding function on group
$G$. If $f$ is smooth, then so is $\tilde f$. Define a map
$$
\eta:\LDiff_{K_{x_{0}}}(G)\rightarrow\LDiff(M)
$$
by the formula
$$
\eta(\square)f=\zeta^{-1}\circ\square\circ\zeta(f),\,f\in
C^{\infty}(M),\,\square\in\LDiff_{K_{x_{0}}}(G).
$$
This map is well defined, since the function
$\square\circ\zeta(f)$ is right-invariant with respect to the
subgroup $K_{x_{0}}$. Evidently, the map $\eta$ is a homomorphism.

Suppose now that $[\mathfrak{p}_{x_{0}},
\mathfrak{k}_{x_{0}}]\subset\mathfrak{p}_{x_{0}}$, so
$\Ad_{K_{x_{0}}} \mathfrak{p}_{x_{0}}\subset\mathfrak{p}_{x_{0}}$.
In some neighborhood of the point $x_{0}\in M$ we can define
coordinates $\{x^{1},\ldots,x^{m}\}$, which correspond to the
point $\pi(\exp(\sum_{i=1}^{m}x^{i}e_{i}))$. The expression of an
operator $\square\in\LDiff(M)$ at the point $x_{0}$ is a
polynomial $P\left(\pd1{x^{1}},\ldots,\pd1{x^{m}}\right)$. Define
a map:
$$
\varkappa:\LDiff(M)\rightarrow S( \mathfrak{p}_{x_{0}}),
$$
by the formula $\varkappa(\square)=P(e_{1},\ldots,e_{m})\in S(
\mathfrak{p}_{x_{0}})$. For any $f\in C^{\infty}(M)$ and $\forall
q\in G$ we have
\begin{eqnarray*}
(\square
f)(qx_{0})=\tau_{q^{-1}}\circ\square(f)(x_{0})=\square\circ\tau_{q^{-1}}(f)(x_{0})=\\
\left[P\left(\pd1{x^{1}},\ldots,\pd1{x^{m}}\right)f\left(\pi\left(q\exp(\sum_{i=1}^{m}
x^{i}e_{i})\right) \right)\right]_{x^{i}=0}=\\
\left[P\left(\pd1{x^{1}},\ldots,\pd1{x^{m}}\right)\tilde
f\left(q\exp(\sum_{i=1}^{m}x^{i}e_{i})\right)\right]_{x^{i}=0}.
\end{eqnarray*}
In particular, for $q\in K_{x_{0}}$ it holds
\begin{eqnarray*}
(\square f)(x_{0})=
\left[P\left(\pd1{x^{1}},\ldots,\pd1{x^{m}}\right)\tilde
f\left(\exp\left(\sum_{i=1}^{m}x^{i}\Ad_{q}e_{i}\right)
\right)\right]_{x^{i}=0}=\nonumber \\ \left[\tilde
P\left(\pd1{x^{1}},\ldots,\pd1{x^{m}}\right)\tilde
f\left(\exp(\sum_{i=1}^{m}x^{i}e_{i})\right)\right]_{x^{i}=0},
\end{eqnarray*}
where $\tilde
P(e_{1},\ldots,e_{m})=P\left(\Ad_{q}e_{1},\ldots,\Ad_{q}e_{m}\right)=
\Ad_{q}P(e_{1},\ldots,e_{m})$, since the map $\varkappa$ does not
depend on a choice of the basis for the space
$\mathfrak{p}_{x_{0}}$ and, in particular, it is not changed by
the transition to the basis $\Ad_{q}e_{i}$. On the other hand,
polynomials $P$ and $\tilde P$ are two expressions of the operator
$\square$ at the point $x_{0}$, so $\tilde
P(e_{1},\ldots,e_{m})=P(e_{1},\ldots,e_{m})$, i.e.
$P(e_{1},\ldots,e_{m})\in I( \mathfrak{p}_{x_{0}})$, where $I(
\mathfrak{p}_{x_{0}})\subset S( \mathfrak{p}_{x_{0}})$ is the set
of $\Ad_{K_{x_{0}}}$ invariants. Note that
$I(\mathfrak{p}_{x_{0}})\subset I(\mathfrak{g})$. Hence we have
the following commutative diagram:
$$
\begin{CD}
I(\mathfrak{p}_{x_{0}})  @>\lambda>> \LDiff_{K_{x_{0}}}(G) \\ @ A
\varkappa AA @ VV\eta V \\ \LDiff(M) @ <\id<< \LDiff(M)
\end{CD}
$$ The structure of the algebra $\LDiff(M)$ was studied in
\cite{Hel1}-\cite{Hel85} with the help of maps $\lambda$ and
$\eta$. We are interested in the representation of a fixed
operator from the algebra $\LDiff(M)$ by a polynomial from the set
$I(\mathfrak{p}_{x_{0}})$. We have constructed the map $\varkappa$
in order to find such a representation. From the definition of the
map $\lambda$ we see that $\eta\circ\lambda\circ\varkappa=\id $
and $\varkappa\circ\eta\circ\lambda=\id $. Hence the maps
$\varkappa,\lambda$ are bijective, the map $\eta$ is surjective,
and the following expansion holds
\begin{equation*}
\LDiff_{K_{x_{0}}}(G)=\lambda(I( \mathfrak{p}_{x_{0}}))\oplus
\ker\eta.
\end{equation*}
Denote by $\LDiff_{ \mathfrak{k}}(G)$ the left ideal in the
algebra $\LDiff(G)$, generated by operators
$X^{l}_{i},\,i=m+1,\ldots,N$ and let
$$
\LDiff_{K_{x_{0}}}^{\mathfrak{k}}(G):=\LDiff_{K_{x_{0}}}(G)
\cap\LDiff_{ \mathfrak{k}}(G).
$$ \begin{Lem}[\cite{Hel85}]
The algebra $\LDiff(G)$ admits the following expansion
\begin{equation*}
\LDiff(G)=\LDiff_{ \mathfrak{k}}(G)\oplus\lambda(S(
\mathfrak{p}_{x_{0}})).
\end{equation*}
\end{Lem}
\begin{theore}[\cite{Hel85}]
If $[ \mathfrak{p}_{x_{0}}, \mathfrak{k}_{x_{0}}]\subset
\mathfrak{p}_{x_{0}}$, then
$\ker\eta=\LDiff_{K_{x_{0}}}^{\mathfrak{k}}(G)$.
\end{theore} \begin{Rem}
\label{Up}We shall use an operator
$\lambda\circ\varkappa(\square)\in \LDiff_{K_{x_{0}}}(G)$ as a
lift $\tilde\square$ of an operator $\square\in\LDiff(M)$ onto the
group $G$.

From the construction of the maps $\lambda,\,\varkappa$ and the
remark \ref{Rem1} we obtain that if
$\left.\square\right|_{x_{0}}=\left.P(X_{1},\ldots,X_{m})\right|_{x_{0}}$,
where $P$ is a polynomial invariant with respect to any
permutation of its arguments, then
$\lambda\circ\varkappa(\square)=P(X_{1}^{l},\ldots,X_{m}^{l})$.
This formula for the lift depends on the expansion $
\mathfrak{g}=\mathfrak{k}_{x_{0}}\oplus\mathfrak{p}_{x_{0}}$.
\end{Rem}  There exists a
unique (up to a constant factor) left- (or right-) invariant
measure on any Lie group (the Haar measure \cite{Hel85},
\cite{Kir}). Denote by $\mu_{G}$ some left-invariant Haar measure
on $G$. A measure on the space $M$, generated by a $G$-invariant
metric is also $G$-invariant. All $G$-invariant measures on $M$
are proportional \cite{Hel85}, and we define such a measure if we
put $\mu_{M}(V)=\mu_{G}(\pi^{-1}(V))$ for any compact set $V\in
M$. The group $K_{x_{0}}$ is compact, so the set $\pi^{-1}(V)$ is
also compact, and $ \mu_{G}(\pi^{-1}(V))<\infty$. The measure
$\mu_{M}$ is left-invariant, since the measure $\mu_{G}$ is
left-invariant.

On all unimodular groups left-invariant measures are also
right-invariant. The change of the point $x_{0}\in M$ to
$x_{1}=qx_{0},q\in G$ leads to the change of the pullback
$\pi^{-1}(V)$ to $\pi^{-1}(V)q^{-1}$, while identifying $M$ with
$G/K_{x_{0}}$. Therefore the $G$-invariant measure $\mu_{M}$ for
the unimodular group $G$ does not depend on the choice of $x_{0}$.

\section{Laplace-Beltrami operator in a moving frame}
\label{LaplaceRepere} \markright{\ref{LaplaceRepere}
Laplace-Beltrami operator} Now we shall find the polynomial $P$
mentioned in the remark \ref{Up} above, corresponding to the
Laplace-Beltrami operator on the space $M$. First, let us obtain
the expression for the Laplace-Beltrami operator in arbitrary
moving frame.

Here we do not regard $M$ as a homogeneous manifold with respect
to the isometry group until the homogeneity is declared
explicitly. Denote the metric on $M$ by $g$. Let
$x^{i},i=1,\dots,m$ be the local coordinates in a domain $U\subset
M$ and $g_{ij}dx^{i}dx^{j}$ be the expression of the metric $g$ on
$U$. The Laplace-Beltrami operator generated by the metric $g$ has
the following form on $U$:
\begin{equation}
\label{Lap}\bigtriangleup_g=\left(\gamma\right)^{-1/2}
\pd1{x^{i}}\left(\sqrt{\gamma}g^{ij}\pd1{x^{j}}\right),
\end{equation}
where $\gamma=|\det g_{ij}|$, and $g^{ij}(x)$ is the inverse of
the matrix $g_{ij}(x)$. The operator $\bigtriangleup$ is conserved
by all isometries of the space $M$. Conversely, if the operator
$\bigtriangleup$ is conserved by some diffeomorphism of the space
$M$, then this diffeomorphism is the isometry \cite{Hel85}.

Let $\xi_{i}=\phi_{i}^{k}(x)\pd1{x^{k}}, i=1,\dots,m$ be vector
fields on $U$ forming a moving frame. Any vector field is a
differential operator of the first order. Using the operation of
composition and nonconstant coefficients we can express any
differential operator on $U$ via this moving frame. The range for
all indices in this section is $1,\dots,m$.

Let $\psi_{j}^{i}$ be the inverse of the matrix $\phi_{i}^{k}$.
Then $\pd1{x^{k}}=\psi^{m}_{k}\xi_{m}$, and $\hat
g_{i,j}:=g(\xi_{i},\xi_{j})=\phi_{i}^{k}\phi_{j}^{m}g_{km}$ are
the coefficients of the metric $g$ with respect to the moving
frame $\xi_{i}$. This implies that $\hat
g^{ij}=\psi_{k}^{i}g^{kn}\psi_{n}^{j}$ and $\hat \gamma
:=|\det\hat g_{ij}|=\phi^{2}\gamma$, where
$\phi=\det\phi_{i}^{k}$. Substituting these formulae in
(\ref{Lap}), we obtain:
\begin{eqnarray*}
\bigtriangleup_g=(\hat\gamma)^{-1/2}\phi\psi_{i}^{q}\xi_{q}\circ\left(\phi^{-1}
\hat\gamma^{1/2}\phi^{i}_{k}\hat
g^{kn}\phi_{n}^{j}\psi_{j}^{p}\xi_{p}\right)=
(\hat\gamma)^{-1/2}\phi\psi_{i}^{q}\xi_{q}\circ \left(\phi^{-1}
\hat\gamma^{1/2}\phi^{i}_{k}\hat g^{kn}\xi_{n}\right)= \\
(\hat\gamma)^{-1/2}\psi_{i}^{q}\phi^{i}_{k}\xi_{q}\circ\left(
\hat\gamma^{1/2}\hat g^{kn}\xi_{n}\right)+
 \phi\psi_{i}^{q}\hat
g^{kn} \xi_{q}\left(\phi^{-1}\phi^{i}_{k}\right)\xi_{n}=
(\hat\gamma)^{-1/2}\xi_{k}\circ\left(\hat\gamma^{1/2}\hat
g^{kn}\xi_{n}\right)+ \hat g^{kn}L_{k}\xi_{n} ,
\end{eqnarray*}
\begin{eqnarray*}\text{where}\quad
L_{k}=\phi\psi_{i}^{q}\xi_{q}\left(\phi^{-1}\phi_{k}^{i}\right) =
\psi_{i}^{q}\xi_{q}\left(\phi_{k}^{i}\right)+
\phi\xi_{k}\left(\phi^{-1}\right)=
\psi_{i}^{q}\xi_{q}\left(\phi_{k}^{i}\right)-
\phi^{-1}\xi_{k}(\phi).
\end{eqnarray*} Denote $\Phi(x)=\|\phi^{i}_{k}(x)\|$. Then using the well known
formula $\det\exp(A)=\exp(\text{Tr}A)$, where $A$ is an arbitrary
complex matrix, we get:
\begin{eqnarray}\label{TrDet}
\phi^{-1}\xi_{k}(\phi)=\xi_{k}(\ln\phi)=
\xi_{k}\left(\ln\exp(\text{Tr}\ln\Phi)\right)=\nonumber\\
\xi_{k}\left(\text{Tr}\ln\Phi\right)=
\text{Tr}\left(\Phi^{-1}\xi_{k}(\Phi)\right)=
\psi^{q}_{i}\xi_{k}(\phi^{i}_{q}),
\end{eqnarray}
since $\text{Tr}(AB)=\text{Tr}(BA)$ for any square matrices $A$
and $B$. On the other hand, the following equations for
commutators of vector fields
\begin{eqnarray*}
\left[\xi_{i},\xi_{j}\right] =
\xi_{i}\left(\phi^{k}_{j}\right)\pd1{x^{k}} -
\xi_{j}\left(\phi^{k}_{i}\right)\pd1{x^{k}}=
\left(\xi_{i}\left(\phi^{k}_{j}\right) -
\xi_{j}\left(\phi^{k}_{i}\right)\right)\psi_{k}^{q}\xi_{q}=:
c^{q}_{ij}\xi_{q}
\end{eqnarray*}
define the functions
$$
c^{q}_{ij}=\left(\xi_{i}\left(\phi^{k}_{j}\right) -
\xi_{j}\left(\phi^{k}_{i}\right)\right)\psi_{k}^{q}
$$
on $U$. So, in view of (\ref{TrDet}), we obtain:
$$
L_{k}=\left(\xi_{q}\left(\phi^{i}_{k}\right) -
\xi_{k}\left(\phi^{i}_{q}\right)\right)\psi_{i}^{q}=c^{q}_{qk}.
$$
Thus we get the formula for the Laplace-Beltrami operator in the
moving frame $\xi_{i}$:
\begin{equation}
\label{LapRep} \bigtriangleup_g=\left(\hat\gamma\right)^{-1/2}
\xi_{q}\circ\left(\sqrt{\hat\gamma}\hat g^{qn}\xi_{n}\right)+\hat
g^{kn}c^{q}_{qk}\xi_{n}.
\end{equation} Let now $\xi_{i}$ be Killing vector fields for the metric $g$ in
$U$. Transform the formula (\ref{LapRep}) to the form
$\bigtriangleup_g=a^{ij}\xi_{i}\circ\xi_{j}+b^{i}\xi_{i}$. It is
clear that $a^{ij}=\hat g^{ij}$ and we only have to find the
coefficients $b^{i}$. The well known formula
$$
X\left(g(Y,Z)\right)=(\pounds_{X}g)(Y,Z)+g([X,Y],Z)+g(Y,[X,Z]),
$$
where $X,Y,Z$ are vector fields on $M$, $\pounds_{X}$ is a Lie
derivative with respect to a field $X$, and formulae
$\pounds_{\xi_{k}}g=0$, (\ref{TrDet}), (\ref{LapRep}) imply
\begin{eqnarray}
\label{calc} b^{i}\hat
g_{ij}=\hat\gamma^{-1/2}\xi_{k}(\hat\gamma^{1/2}\hat g^{ki})\hat
g_{ij}+\hat g^{ki}c_{qk}^{q}\hat g_{ij}= \xi_{k}(\hat g^{ki}\hat
g_{ij})- \xi_{k}(\hat g_{ij})\hat g^{ki}+
\frac{1}{2\hat\gamma}\xi_{k}(\hat\gamma)\hat g^{ki}\hat g_{ij}
\nonumber \\ +c_{qk}^{q}\delta^{k}_{j}=
\xi_{k}(\delta_{j}^{k})-\xi_{k}( g(\xi_{i},\xi_{j}))\hat
g^{ki}+\frac{1}{2}\hat g^{qi}\xi_{k}(\hat
g_{qi})\delta_{j}^{k}+c_{qj}^{q}=
-g([\xi_{k},\xi_{i}],\xi_{j})\hat g^{ki}\nonumber \\ -
g(\xi_{i},[\xi_{k},\xi_{j}])\hat g^{ki}+ \frac{1}{2}\hat g^{qi}
g([\xi_{j},\xi_{q}],\xi_{i})+ \frac12\hat
g^{qi}g(\xi_{q},[\xi_{j},\xi_{i}])+ c_{qj}^{q}=\nonumber \\ -\hat
g_{iq}c_{kj}^{q}\hat g^{ki}+ \frac{1}{2}\hat g^{qi}c_{jq}^{k}\hat
g_{ki}+ \frac{1}{2}\hat g^{qi}c_{ji}^{k}\hat g_{qk}+c^{q}_{qj}=
-c_{qj}^{q}+\frac12c_{jq}^{q}+\frac12c_{jq}^{q}+
c_{qj}^{q}=c_{jq}^{q},
\end{eqnarray}
taking into account the antisymmetry of the tensor $c^{q}_{ki}$
with respect to lower indices. Thus we obtain
$b^{i}=c_{jq}^{q}\hat g^{ji}$. We can summarize this reasoning in
the following proposition:
\begin{proposit}
\label{LemUp} In the moving frame $\xi_{i}$ the Laplace-Beltrami
operator have the following form
\begin{equation*}
\bigtriangleup_{g}=\hat g^{ij}\xi_{i}\circ\xi_{j}+c_{jq}^{q}\hat
g^{ji}\xi_{i}.
\end{equation*}
If the space $M$ is homogeneous and $\xi_{i}=X_{i}$ in notations
of section \ref{InvOperators}, then the remark \ref{Up} implies
that the lift of the operator $\bigtriangleup_{g}$ onto the group
$G$ has the form:
\begin{equation*}
\tilde\bigtriangleup_{g}=\left.\hat
g^{ij}\right|_{x_{0}}X_{i}^{l}\circ X_{j}^{l}+\left.c_{jq}^{q}\hat
g^{ji}\right|_{x_{0}}X_{i}^{l}.
\end{equation*}
\end{proposit}
\begin{Rem}
Sometimes vector fields $\xi_{i}$ can be chosen in such a way that
$c_{jq}^{q}=0$. In this case we have $\bigtriangleup_{g}=\hat
g^{ij}\xi_{i}\circ\xi_{j}$ and
$\tilde\bigtriangleup_{g}=\left.\hat
g^{ij}\right|_{x_{0}}X_{i}^{l}\circ X_{j}^{l}$.
\end{Rem} In the sequel, we shall use the expression for the Riemannian
connection $\nabla$ in the moving frame $\xi_{i}$ given by the
following lemma:
\begin{Lem}[\cite{Besse}, 7.27]
\begin{equation*}
g(\nabla_{\xi_{i}}\xi_{j},\xi_{k})=\frac12g(\xi_{i},[\xi_{j},\xi_{k}])+
\frac12g(\xi_{j},[\xi_{i},\xi_{k}])+\frac12g([\xi_{i},\xi_{j}],\xi_{k}).
\end{equation*}
In particular, for $i=j$
\begin{equation}\label{conpart}
g(\nabla_{\xi_{i}}\xi_{i},\xi_{k})=g(\xi_{i},[\xi_{i},\xi_{k}]),
\end{equation}
without summation over index $i$.
\end{Lem} \section{Two-point homogeneous Riemannian spaces}
\label{twopoint} \markright{\ref{twopoint} Two point homogeneous
Riemannian spaces}  The main characteristic for the system of two
classical particles is the distance between them. If the
configuration space $Q$ is homogeneous and isotropic, this
distance is the only geometric invariant for $Q$. These spaces are
called {\it two-point homogeneous spaces} \cite{Wolf}, i.e. any
pair of points on such space can be transformed by means of
appropriate isometry to any other pair of points with the same
distance between them. In the following, $Q$ denotes the two-point
homogeneous connected Riemannian space. The classification of
these spaces can been found in \cite{Tits}, \cite{Wang}, (see also
\cite{Matsumoto}, \cite{Wolf}), and is as follows:
\begin{enumerate}
\item the Euclidean space $\mathbf{ E}^{n}$;
\item the sphere $\mathbf{ S}^{n}$;
\item the real projective space $\mathbf{P}^{n}(\mathbb{R})$;
\item the complex projective space $\mathbf{P}^{n}(\mathbb{C})$;
\item the quaternion projective space $\mathbf{P}^{n}(\mathbb{H})$;
\item the Cayley projective plane $\mathbf{P}^{2}(\mathbb{C}a)$;
\item the real hyperbolic space (Lobachevski space) $\mathbf{H}^{n}(\mathbb{R})$;
\item the complex hyperbolic space $\mathbf{H}^{n}(\mathbb{C})$;
\item the quaternion hyperbolic space  $\mathbf{H}^{n}(\mathbb{H})$;
\item the Cayley hyperbolic plane $\mathbf{H}^{2}(\mathbb{C}a)$.
\end{enumerate} There are different equivalent approaches to classification of
these spaces.  Recall that the {\it rank of a symmetric space} is
the dimension of its maximal flat completely geodesic submanifold.
\begin{theore}
Let $Q$ be a connected Riemannian space, $G$ is the identity
component of the isometry group for $M$, and $I_{x}$ is a
stationary subgroup for a point $x$. Then the following conditions
\ref{q1}, \ref{q2}, \ref{q3} are equivalent
\begin{enumerate}
\item $Q$ is two point homogeneous;\label{q1}
\item the action of the stationary subgroup $I_{x}$ on all unit spheres $T_{x}Q,\;\forall x\in Q$ in the tangent
spaces is transitive; in other words, $Q$ is isotropic; \label{q2}
\item $Q$ is the symmetric space of the rank one. \label{q3}
\end{enumerate}
If any of these condition is satisfied, then all geodesics on the
compact space $Q$ are closed and have the same length. This
follows from the homogeneity and isotropy of $Q$.
\end{theore}
This result has been proved in \cite{Wolf}, lemma 8.12.1,
\cite{Wang}, \cite{Tits}, see also references in \cite{Hel78},
p.~535.

Let now $Q$ be the two-point homogeneous compact Riemannian space
(i.e. the space of one of the types 2-6). We assume that the point
$x_{0}$ is fixed (the index $x_{0}$ will sometimes be omitted in
the following). All geodesics on $Q$ are closed and have the same
length equal $2\diam Q$, where $\diam Q$ is the maximal distance
between two points of the space $Q$. Put $R=2\diam Q/\pi$ for the
space $\mathbf{P}^{n}(\mathbb{R})$ and $R=\diam Q/\pi$ for the
other compact two-point homogeneous Riemannian spaces. The maximal
sectional curvature of all these spaces is $R^{-2}$ and the
minimal  sectional curvature of the spaces
$\mathbf{P}^{n}(\mathbb{C}),\;\mathbf{P}^{n}(\mathbb{H}),\;
\mathbf{P}^{2}(\mathbb{C}a)$ is $(2R)^{-2}$.

Let $\tilde\gamma(s),\; s\in [0,\diam Q)$ be some geodesic, which
is parameterized by a natural parameter $s$ and
$\tilde\gamma(0)=x_{0}$. Since the space $Q$ is symmetric, in the
algebra $\mathfrak{g}$ there exists a complementary subspace
$\mathfrak{p}$ with respect to the subalgebra $\mathfrak{k}$ such
that
$[\mathfrak{k},\mathfrak{p}]\subset\mathfrak{p},\;[\mathfrak{p},
\mathfrak{p}]\subset\mathfrak{k}$. The space $ \mathfrak{p}$ can
be naturally identified with the space $T_{x_{0}}Q$. Under this
identification the restriction of the Killing form for the algebra
$\mathfrak{g}$ onto the space $\mathfrak{p}$ and the scalar
product on $T_{x_{0}}Q$ are proportional. In particular, the
decomposition $\mathfrak{g}=\mathfrak{p}\oplus\mathfrak{k}$ is
uniquely determined by the point $x_{0}$.\label{decompose} Let
$(\cdot,\cdot)$ be the scalar product on the algebra
$\mathfrak{g}$ such that it is proportional to the Killing form
and its restriction onto the subspace $ \mathfrak{p}\cong
T_{x_{0}}Q$ coincides with the Riemannian metric $g$ on
$T_{x_{0}}Q$. The inclusions
$$
[\mathfrak{p},[\mathfrak{k},\mathfrak{p}]]\subset\mathfrak{k},\;[\mathfrak{p},
[\mathfrak{k},\mathfrak{k}]]\subset\mathfrak{p}
$$
and the definition of the Killing form imply that the spaces
$\mathfrak{p}$ and $ \mathfrak{k}$ are orthogonal to each other
with respect to the scalar product $(\cdot,\cdot)$. From the
results of \cite{Hel78}, \cite{Loos} we can extract the following
proposition.
\begin{proposit}\label{prop1}
The algebra $\mathfrak{g}$ admits the following expansion into the
direct sum of subspaces:
\begin{equation*}
\mathfrak{g}=\mathfrak{a}\oplus\mathfrak{k}_{0}\oplus\mathfrak{k}_{\lambda}\oplus
\mathfrak{k}_{2\lambda}\oplus\mathfrak{p}_{\lambda}\oplus
\mathfrak{p}_{2\lambda},
\end{equation*}
where $\dim\mathfrak{a}=1$, $\lambda$ is a nontrivial linear form
on the space $ \mathfrak{a}$,
$\dim\mathfrak{k}_{\lambda}=\dim\mathfrak{p}_{\lambda}=q_{1}$,
$\dim\mathfrak{k}_{2\lambda}=\dim\mathfrak{p}_{2\lambda}=q_{2}$,
$\mathfrak{p}=\mathfrak{a}\oplus\mathfrak{p}_{\lambda}\oplus\mathfrak{p}_{2\lambda}$,
$\mathfrak{k}=\mathfrak{k}_{0}\oplus\mathfrak{k}_{\lambda}\oplus\mathfrak{k}_{2\lambda}$;
where $q_{1},q_{2}\in \{0\}\cup \mathbb{N}$, the subalgebra
$\mathfrak{a}$ is the maximal commutative subalgebra in the
subspace $\mathfrak{p}$ and it corresponds to the tangent vectors
to the geodesic $\tilde\gamma$ at the point $x_{0}$. Besides, the
following inclusions are valid:
\begin{align}\label{inclusions}
[\mathfrak{a},\mathfrak{p}_{\lambda}]\subset\mathfrak{k}_{\lambda},\;
[\mathfrak{a},\mathfrak{f}_{\lambda}]\subset\mathfrak{p}_{\lambda},\;
[\mathfrak{a},\mathfrak{p}_{2\lambda}]\subset\mathfrak{k}_{2\lambda},\;
[\mathfrak{a},\mathfrak{f}_{2\lambda}]\subset\mathfrak{p}_{2\lambda},\;
[\mathfrak{a},\mathfrak{k}_{0}]=0,\;
[\mathfrak{k}_{\lambda},\mathfrak{p}_{\lambda}]\subset\mathfrak{p}_{2\lambda}
\oplus\mathfrak{a},\nonumber \\
[\mathfrak{k}_{\lambda},\mathfrak{k}_{\lambda}]\subset\mathfrak{k}_{2\lambda}\oplus
\mathfrak{k}_{0},\;
[\mathfrak{p}_{\lambda},\mathfrak{p}_{\lambda}]\subset\mathfrak{k}_{2\lambda}\oplus
\mathfrak{k}_{0},\;
[\mathfrak{k}_{2\lambda},\mathfrak{k}_{2\lambda}]\subset\mathfrak{k}_{0},\;
[\mathfrak{p}_{2\lambda},\mathfrak{p}_{2\lambda}]\subset\mathfrak{k}_{0},\;
[\mathfrak{k}_{2\lambda},\mathfrak{p}_{2\lambda}]\subset\mathfrak{a},\nonumber\\
[\mathfrak{k}_{\lambda},\mathfrak{k}_{2\lambda}]\subset\mathfrak{k}_{\lambda},\;
[\mathfrak{k}_{\lambda},\mathfrak{p}_{2\lambda}]\subset\mathfrak{p}_{\lambda},\;
[\mathfrak{p}_{\lambda},\mathfrak{k}_{2\lambda}]\subset\mathfrak{p}_{\lambda},\;
[\mathfrak{p}_{\lambda},\mathfrak{p}_{2\lambda}]\subset\mathfrak{k}_{\lambda}.
\end{align} Moreover, for any basis $e_{\lambda,i},\;i=1,\dots,q_{1}$ in the
space $\mathfrak{p}_{\lambda}$ and any basis
$e_{2\lambda,i},\;i=1,\dots,q_{2}$ in the space
$\mathfrak{p}_{2\lambda}$ there are the basis
$f_{\lambda,i},\;i=1,\dots,q_{1}$ in the space
$\mathfrak{k}_{\lambda}$ and the basis
$f_{2\lambda,i},\;i=1,\dots,q_{2}$ in the space
$\mathfrak{k}_{2\lambda}$ such that:
\begin{align}
\label{commSpesial}
[Z,e_{\lambda,i}]=-\lambda(Z)f_{\lambda,i},\;[Z,f_{\lambda,i}]=\lambda(Z)e_{\lambda,i},\;
i=1,\dots,q_{1},\nonumber\\
[Z,e_{2\lambda,i}]=-2\lambda(Z)f_{2\lambda,i},\;[Z,f_{2\lambda,i}]=2\lambda(Z)e_{2\lambda,i},\;
i=1,\dots,q_{2},\forall Z\in\mathfrak{a}.
\end{align}
If a vector $\Lambda\in\mathfrak{a}$ satisfy the condition
$(\Lambda,\Lambda)=R^{2}$, then $|\lambda(\Lambda)|=\dfrac12$.
\end{proposit} Nonnegative integers $q_{1}$ and $q_{2}$ are said to be {\it
multiplicities of the space} $Q$. They characterize $Q$ uniquely
up to the exchange
$\mathbf{S}^{n}\leftrightarrow\mathbf{P}^{n}(\mathbb{R})$. For the
spaces $\mathbf{S}^{n}$ and $\mathbf{P}^{n}(\mathbb{R})$ we have
$q_{1}=0,\;q_{2}=n-1$; for the space
$\mathbf{P}^{n}(\mathbb{C}):\; q_{1}=2n-2,\;q_{2}=1$; for the
space $\mathbf{P}^{n}(\mathbb{H}):\; q_{1}=4n-4,\;q_{2}=3$; and
for the space $\mathbf{P}^{2}(\mathbb{C}a):\; q_{1}=8,\;q_{2}=7$.
Conversely, for the spaces $\mathbf{S}^{n}$ and
$\mathbf{P}^{n}(\mathbb{R})$ we could reckon that
$q_{1}=n-1,q_{2}=0$. Our choice corresponds to the isometries
$\mathbf{P}^{1}(\mathbb{C})\cong\mathbf{S}^{2},\,
\mathbf{P}^{1}(\mathbb{H})\cong\mathbf{S}^{4}$.
\begin{Rem}\label{GeodesicSubmanifolds}
The space $\mathfrak{a}\oplus\mathfrak{p}_{2\lambda}$ generates in
the space $Q$ a completely geodesic submanifold  of the constant
sectional curvature $R^{-2}$ and dimension $q_{2}+1$. For all the
above mentioned spaces $Q$ with the exception of ${\bf
P}^{n}(\mathbb{R})$ this submanifold is a sphere. For the space
${\bf P}^{n}(\mathbb{R})$ this submanifold is the space ${\bf
P}^{q_{2}+1}(\mathbb{R})$. If $q_{1}\neq 0$, the element $\Lambda$
and an arbitrary nonzero element from the space
$\mathfrak{p}_{\lambda}$ generate in $Q$ a completely geodesic two
dimensional submanifolds of the constant curvature $R^{-2}$.
\end{Rem}
Choose a vector $\Lambda\in\mathfrak{a}$ such that
$\lambda(\Lambda)=\frac12$. The following proposition easily
follows from the proposition \ref{prop1}:
\begin{proposit}
\label{prop2} The spaces $\mathfrak{a}\oplus\mathfrak{k}_{0},\;
\mathfrak{k}_{\lambda}\oplus\mathfrak{p}_{\lambda},\;
\mathfrak{k}_{2\lambda}\oplus\mathfrak{p}_{2\lambda}$ are pairwise
orthogonal. One has
\begin{align}
\label{scal}
(e_{\lambda,i},e_{\lambda,j})=(f_{\lambda,i},f_{\lambda,j}),\;
(e_{\lambda,i},f_{\lambda,j})=-(f_{\lambda,i},e_{\lambda,j}),\;i,j=1,\dots,q_{1},\nonumber\\
(e_{2\lambda,i},e_{2\lambda,j})=(f_{2\lambda,i},f_{2\lambda,j}),\;
(e_{2\lambda,i},f_{2\lambda,j})=-(f_{2\lambda,i},e_{2\lambda,j}),\;i,j=1,\dots,q_{2}.
\end{align}
 In particular,
\begin{equation*}
(e_{\lambda,i},f_{\lambda,i})=0,\;i=1,\dots,q_{1},\;(e_{2\lambda,j},f_{2\lambda,j})=0,\;
j=1,\dots,q_{2}.
\end{equation*}
\end{proposit}
\begin{proof}
The Jacobi  identity and the $\Ad_{G}$-invariance of the metric
$(\cdot,\cdot)$ imply that the operator $T_{\Lambda}:X
\rightarrow[\Lambda,[\Lambda,X]]$ is symmetric on the space
$\mathfrak{g}$. This operator has the following eigenspaces
$\mathfrak{a}\oplus\mathfrak{k}_{0},\;
\mathfrak{k}_{\lambda}\oplus\mathfrak{p}_{\lambda},\;
\mathfrak{k}_{2\lambda}\oplus\mathfrak{p}_{2\lambda}$ with
eigenvalues
$0,\;-\lambda^{2}(\Lambda)=-\dfrac14,-4\lambda^{2}(\Lambda)=-1$,
respectively. Thus, these eigenspaces are orthogonal to each
other. Let us prove the first equality from (\ref{scal}). The
$\Ad_{G}$-invariance of the metric $(\cdot,\cdot)$ and the
equality (\ref{commSpesial}) give
$$
\lambda(\Lambda)(e_{\lambda,i},e_{\lambda,j})=([\Lambda,f_{\lambda,i}],e_{\lambda,j})=
-(f_{\lambda,i},[\Lambda,e_{\lambda,j}])=\lambda(\Lambda)(f_{\lambda,i},f_{\lambda,j}).
$$
Similar calculations prove other equalities from (\ref{scal}).
\end{proof} The Jacobi identity and formulae (\ref{commSpesial}) give
$[Z,[e_{\lambda,i},f_{\lambda,i}]]=0$. Thus the relations
(\ref{inclusions}) give
$[e_{\lambda,i},f_{\lambda,i}]\in\mathfrak{a}$. Let
$[e_{\lambda,i},f_{\lambda,i}]=:\varkappa_{i}\Lambda$. The
$\Ad_{G}$-invariance of the metric $(\cdot,\cdot)$ leads to
$$
0=(\Lambda,[e_{\lambda,i},f_{\lambda,i}])+([e_{\lambda,i},\Lambda],f_{\lambda,i})=
\varkappa_{i}(\Lambda,\Lambda)+\lambda(\Lambda)(f_{\lambda,i},f_{\lambda,i}),
$$
and
$$
\varkappa_{i}=-\frac{\lambda(\Lambda)}{(\Lambda,\Lambda)}(f_{\lambda,i},f_{\lambda,i})=
-\frac{\lambda(\Lambda)}{(\Lambda,\Lambda)}(e_{\lambda,i},e_{\lambda,i}).
$$
Similarly, we get:
$$
[e_{2\lambda,i},f_{2\lambda,i}]=-\frac{2\lambda(\Lambda)}{(\Lambda,\Lambda)}
(f_{2\lambda,i},f_{2\lambda,i})\Lambda=
-\frac{2\lambda(\Lambda)}{(\Lambda,\Lambda)}(e_{2\lambda,i},e_{2\lambda,i})\Lambda.
$$ Using the freedom provided by the proposition \ref{prop1},
we choose the bases $\{e_{\lambda,i}\}_{i=1}^{q_{1}}$ in the space
$\mathfrak{p}_{\lambda}$ and $\{e_{2\lambda,j}\}_{j=1}^{q_{2}}$ in
the space $\mathfrak{p}_{2\lambda}$ to be orthogonal with norms of
all their elements equal $R$. Thus, the elements
$\Lambda,e_{\lambda,i},e_{2\lambda,j},
i=1,\dots,q_{1},j=1,\dots,q_{2}$ form the orthogonal basis in the
space $\mathfrak{p}$ and the elements
$f_{\lambda,i},\;f_{2\lambda,j},i=1,\dots,q_{1},j=1,\dots,q_{2}$
form the orthogonal basis in the space
$\mathfrak{k}_{\lambda}\oplus\mathfrak{k}_{2\lambda}$, due to the
proposition \ref{prop2}. Note that
$(f_{\lambda,i},f_{\lambda,i})=R^{2},i=1,\dots,q_{1},\;
(f_{2\lambda,j},f_{2\lambda,j})=R^{2},j=1,\dots,q_{2}$.
\begin{proposit} \label{prop3}
\begin{enumerate}
\item The relations (\ref{commSpesial}) can be rewritten in the
following form:
\begin{align}\label{commSpesial1}
[\Lambda,e_{\lambda,i}]=-\frac12f_{\lambda,i},\;[\Lambda,f_{\lambda,i}]=\frac12e_{\lambda,i},\;
[e_{\lambda,i},f_{\lambda,i}]=-\frac12\Lambda,\nonumber\\
(e_{\lambda,i},e_{\lambda,j})=(f_{\lambda,i},f_{\lambda,j})=\delta_{ij}R^{2},\;
i,j=1,\dots,q_{1},\nonumber\\
[\Lambda,e_{2\lambda,i}]=-f_{2\lambda,i},\;[\Lambda,f_{2\lambda,i}]=
e_{2\lambda,i},\;
[e_{2\lambda,i},f_{2\lambda,i}]=-\Lambda,\nonumber\\
(e_{2\lambda,i},e_{2\lambda,j})=(f_{2\lambda,i},f_{2\lambda,j})=\delta_{ij}R^{2},\;i,j=1,\dots,q_{2},
\;(\Lambda,\Lambda)=R^{2}.
\end{align} \item Let $X$ and $Y$ be some elements from the basis
\begin{equation}\label{basism}
\Lambda,e_{\lambda,i},f_{\lambda,i},e_{2\lambda,j},f_{2\lambda,j},\;i=1,\dots,q_{1},
j=1,\dots,q_{2}
\end{equation}
of the space $
\mathfrak{m}:=\mathfrak{a}\oplus\mathfrak{k}_{\lambda}\oplus\mathfrak{k}_{2\lambda}
\oplus\mathfrak{p}_{\lambda}\oplus\mathfrak{p}_{2\lambda}$. Let
$X'_{\mathfrak{m}}$ be the projection of an element
$X'\in\mathfrak{g}$ on the space $\mathfrak{m}$ with respect to
the expansion $\mathfrak{g}=\mathfrak{k}_{0}\oplus\mathfrak{m}$.
Expand the element $[X,Y]_{\mathfrak{m}}$ in the basis
(\ref{basism}). Then its coordinates with respect to the elements
$X,Y$ are equal to zero.
\end{enumerate}
\end{proposit}
\begin{proof}
The relations (\ref{commSpesial1}) are evident. In view of the
inclusions from the proposition \ref{prop1} it is sufficient to
prove the second statement only in the following cases: a)
$X=e_{\lambda,i},\;Y=f_{2\lambda,j}$ and b)
$X=f_{\lambda,i},\;Y=f_{2\lambda,j}$. Consider the case a). From
(\ref{inclusions}) we get
$[f_{2\lambda,j},e_{\lambda,i}]\in\mathfrak{p}_{\lambda}$. The
$\Ad_{G}$-invariance of the metric $(\cdot,\cdot)$ gives
$\left([f_{2\lambda,j},e_{\lambda,i}],e_{\lambda,i}\right)=-
\left(e_{\lambda,i},[f_{2\lambda,j},e_{\lambda,i}]\right),\;i=1,\dots,q_{1},j=1,\dots,q_{2}$
and then $[f_{2\lambda,j},e_{\lambda,i}]\bot e_{\lambda,i}$. Now,
taking into account the orthogonality of the basis
$\{e_{\lambda,i}\}_{i=1}^{q_{1}}$ of the space
$\mathfrak{p}_{\lambda}$, we obtain the second statement in the
case a). The case b) is completely similar.
\end{proof}

\section{Homogeneous submanifolds of two-body problem on two-point
homogeneous compact Riemannian spaces}
\label{hom}\markright{\ref{hom} Homogeneous submanifolds} Let an
operator $H$ be as in the section \ref{Introduction};
$\pi_{i},\,i=1,2$ is the projection on the $i$-th factor in the
decomposition of $M=Q\times Q,\,\dim_{\mathbb{R}}Q=n$, and
$\rho(x_{1},x_{2})$ the distance between the points $x_{1},
x_{2}\in Q$. The function
$\rho_{2}(x):=\rho(\pi_{1}(x)),\pi_{2}(x)), x\in M$ determines the
distance between particles. The free Hamiltonian $H_{0}$ for the
system of two particles on the space $M$ is the Laplace-Beltrami
operator for the metric
$$
g_{2}:=m_{1}\pi^{*}_{1}g + m_{2}\pi^{*}_{2}g
$$
on this space, multiplied by $-1/2$, where $\pi^{*}_{i}g$ is the
pullback of the metric $g$ with respect to the projection on the
$i$-th factor. In order to find the explicitly invariant
expression for the operator $H_{0}$, consider the foliation of the
space $M$ by submanifolds $F_{p}$ that correspond to the constant
level of the function $\rho_{2}$. The layer $F_{p}\subset M$ is
$G$-homogeneous Riemannian manifolds with respect to the
restriction of the metric $g_{2}$ on it; therefore, we can use the
construction from the section \ref{InvOperators}. To "glue" these
constructions for different $p$, we shall do the following. Choose
a smooth curve $\gamma(p)\subset M$ to be transversal to the
layers $F_{p},\,p>0$ and identify each $F_{p}$ with the factor
spaces $G/K_{p}$, where $K_{p}$ is a stationary subgroup for the
point $\gamma_{p}$. Let $\mathfrak{k}_{p}$ be a Lie algebra for
$K_{p}$. Note that the layer $F_{0}$ is diffeomorphic to the space
$Q$. Assume that the following condition is valid.
\begin{Con}
\label{A} A function $p\rightarrow x_{p}$ on some interval
$I\subset\mathbb{R}_{+}$ is a regular parametrization for the
curve $\gamma$ in $M$. This curve intersects each $F_{p},\,p>0$
once, and the set $M':=\bigcup\limits_{p\in I}F_{p}$ is a
connected dense open submanifold in $M$. Stationary subgroups for
the points $x_{p}$ coincide as $p\in I$.
\end{Con}

For two-point homogeneous compact Riemannian spaces $Q$ the curve
from the condition \ref{A} can be chosen in the following way. Let
$\tilde\gamma(t):(a,b)\subset\mathbb{R}\rightarrow Q$ be some
geodesic on the space $Q$, where $t$ is a natural parameter,
$0\in(a,b)$. The geodesic $\tilde\gamma$ realizes the strong
minimum for length of curves between any two points on
$\tilde\gamma$, if they are sufficiently close to each other
(\cite{KoNo1}). Let $\tilde\gamma(t_{1})$ and
$\tilde\gamma(t_{2})$ be  two such points. Denote by $\Gamma$ the
segment of the geodesic $\gamma$ between those points.

Let $K$ be a subgroup of the group $G$ consisting of all
transformations conserving the points $\tilde\gamma(t_{1})$ and
$\tilde\gamma(t_{2})$. Any isometry transforms a geodesic into a
geodesic, so any element $q\in K$ conserves the segment $\Gamma$
and consequently $q$ conserves the whole geodesic $\tilde\gamma$.
Thus, any point of $\Gamma$ is a fixed point with respect to $q$,
otherwise its motions along $\Gamma$ would lead to changing the
distance from this point to the ends of $\Gamma$. Consider the
maximal interval $\Gamma'\supset\Gamma$, consisting of fixed
points of $K$. The continuity of the $K$-action on the space $Q$
implies that $\Gamma'$ is closed. On the other hand, the group $K$
conserves the geodesic $\tilde\gamma$, so it conserves those
points on $\tilde\gamma$ near the ends of $\Gamma'$ for which the
distance from the ends of $\Gamma'$ is realized only by
$\tilde\gamma$. Thus the interval $\Gamma'$ is open (as a subset
$\tilde\gamma$) and coincides with $\tilde\gamma$, since the
latter is connected.

Any geodesic is uniquely defined by any pair of its points, if
they are close enough, therefore the group $G$ acts transitively
on the set of all geodesics of the space $Q$.

Let $s_{1},s_{2}:[0,\diam Q)\rightarrow (a,b)$ be smooth
functions, $s_{1}$ is decreasing, $s_{2}$ is increasing,
$s_{1}(0)=s_{2}(0)=0$, and
$\rho(\tilde\gamma(s_{1}(p)),\tilde\gamma(s_{2}(p)))\equiv
p,\,p\in[0,\diam Q),\,s_{1}'(p)^{2}+s_{2}'(p)^{2}\neq 0$. Define a
curve $\gamma:[0,\diam Q)\rightarrow M$ by the formula
$\gamma(p)=(\tilde\gamma(s_{1}(p)),\tilde\gamma(s_{2}(p)))\in M$.
Below we shall formulate the condition \ref{B}, which implies that
the stationary subgroup of the group $G$, corresponding to the
points $\gamma(p),\,p\in(0,\diam Q)$ and as shown above containing
$K$, equals $K$. The validity of this condition will be verified
later in the section \ref{generaltwobody}. Obviously, the other
requirements of the condition \ref{A} are realized for $I=(0,\diam
Q)$.

In this case we can identify the manifold $M'$ with the space
$I\times G/K$, where $K_{x_{p}}\equiv K,\,p\in I$, by the
following formula:
$$
I\times\left(G/K\right)\ni (p,bK)\longleftrightarrow bx_{p}\in M'.
$$
Let $\mu_{2}$ be a measure on $M$, generated by the metric
$g_{2}$. Using the identification above, carry this measure on the
space $I\times\left(G/K\right)$, saving for it the same notation
$\mu_{2}$. Since the difference $M\backslash M'$ has a zero
measure, we get the following isomorphism between spaces of
measurable square integrable functions:
$$
\mathcal{L}^{2}\left(M,\mu_{2}\right)\cong\mathcal{L}^{2}\left(I\times
\left(G/K\right),\mu_{2}\right).
$$
In the following, for simplicity it will be convenient to change
the parametrization of the interval $I$ using some function
$p(r),\,p'(r)\neq 0,\,r\in I'\subset\mathbb{R}_{+}$. In this case
we will write $F_{r}:=F_{p(r)}$. Since the group $G$ acts only on
the second factor of the space $M'=I\times\left(G/K\right)$, we
can generalize the construction for the lift of differential
operators from the section \ref{InvOperators} and find for a
$G$-invariant differential operator on the space
$I\times\left(G/K\right)$ its lift onto the space $I\times G$.

Let $\mathfrak{p}$ be a subspace in $ \mathfrak{g}$, complimentary
to the subalgebra $ \mathfrak{k}\equiv\mathfrak{k}_{p}$ such that
$[ \mathfrak{p}, \mathfrak{k}]\subset \mathfrak{p}$. Let
$e_{1},\ldots e_{2n-1}$ be a basis in $ \mathfrak{p},\,
X_{1},\ldots,X_{2n-1}$ the corresponding Killing vector fields on
the space $M'$, and $ X_{i}^{l},X_{i}^{r}$ the corresponding left-
and right-invariant vector fields on the group $G$. Define a
vector field tangent to the curve $\gamma$ by the formula
$X_{0}=\frac{d}{dr}x_{p(r)}$. Since
$$
dL_{q}X_{0}=\frac{d}{dr}L_{q}x_{p(r)}=\frac{d}{dr}x_{p(r)}=X_{0},\,\forall
q\in K,
$$
it is possible to spread the vector $X_{0}$ by left shifts to the
whole space $M'$ and obtain the smooth vector field on $M'$ with
the same notation $X_{0}$. The fields $X_{i},\,i=0,\ldots,2n-1$
form the moving frame in some neighborhood of the curve
$\gamma(p),\,p\in (0,\diam Q)$, if the matrix $\Gamma$, consisting
of the pairwise scalar products of the fields $X_{i}$, is
nondegenerate on $\gamma(p),\,p\in (0,\diam Q)$. Besides, at those
points of the curve $\gamma$, where $\det\Gamma\neq 0$, the
stationary subgroup of the group $G$, containing, as shown above,
the group $K$, coincides with $K$, in view of the decomposition
$\mathfrak{g}=\mathfrak{p}\oplus\mathfrak{k}$. The next condition
will be verified later in section \ref{generaltwobody}.
\begin{Con}
\label{B} The matrix $\Gamma$ is nonsingular on the curve
$\gamma(p),\,p\in (0,\diam Q)$.
\end{Con} Express the operator $\bigtriangleup_{g_{2}}$ via the moving frame
$X_{i},\,i=0,\ldots,2n-1$ by the formula (\ref{LapRep}), assuming
$\xi_{i}=X_{i},\,i=0,\ldots,2n-1$, and transform the result to the
form $a^{ij} X_{i}\circ X_{j}+b^{i}X_{i}$. Since the field
$X_{0}$, in contrast to other fields $X_{i}$, is not a Killing
one, after calculations similar to (\ref{calc}), we obtain the
following additional terms:
\begin{equation*}
-\left(\pounds_{X_{0}}g_{2}\right)(X_{i},X_{j})\hat
g_{2}^{0i}+\frac12\hat
g_{2}^{ki}\left(\pounds_{X_{0}}g_{2}\right)(X_{k},X_{i})\delta_{j}^{0},
\end{equation*}
where $\hat g_{2,ij}:=g_{2}(X_{i},X_{j}),\,0\leq i,j\leq 2n-1 $
are components of the metric $g_{2}$ with respect to the moving
frame $X_{i}$. Taking into account $[X_{0},X_{i}]=0,\,\forall
i=0,\ldots,2n-1$, we get:
\begin{equation*}
\left(\pounds_{X_{0}}g_{2}\right)(X_{i},X_{j})=X_{0}g_{2}(X_{i},X_{j})=
X_{0}\left(\hat g_{2,ij}\right).
\end{equation*} Thus, using formula (\ref{TrDet}), we obtain the following
additional term in the formula for the operator:
\begin{eqnarray*}
\frac12X_{0}\left(\hat g_{2,kj}\right)\hat g_{2}^{kj}\hat
g_{2}^{0i}X_{i}-X_{0}\left(\hat g_{2,kj}\right)\hat g_{2}^{0k}\hat
g_{2}^{ji}X_{i}=\frac1{2\hat\gamma}X_{0}(\hat\gamma)\hat
g_{2}^{0i}X_{i}+X_{0}(\hat g^{0i})X_{i}=\nonumber \\
\frac1{\sqrt{\hat\gamma}}X_{0}\left(\sqrt{\hat\gamma}\hat
g_{2}^{0i}\right)X_{i},
\end{eqnarray*}
where $\hat\gamma=\det\hat g_{2,ij}$. Finally, we get:
\begin{equation*}
\bigtriangleup_{g_{2}}=\hat g^{ij}X_{i}\circ X_{j}+c_{jq}^{q}\hat
g^{ji}X_{i}+\frac1{\sqrt{\hat\gamma}}X_{0}\left(\sqrt{\hat\gamma}\hat
g_{2}^{0i}\right)X_{i}.
\end{equation*} The field $X_{0}$ on the space $I'\times\left(G/K\right)$ has the
form $\partial/\partial r$ and its lift on the space $I'\times G$
is tautological. This lift changes only the fields
$X_{i},\,i=1,\ldots,2n-1$. According to the remark \ref{Up} and
lemma \ref{LemUp} we obtain the expression for the lift of the
operator $\bigtriangleup_{g_{2}}$:
\begin{equation}
\label{LapUp} \tilde\bigtriangleup_{g_{2}}=\left.\hat
g^{ij}\right|_{x_{0}}X_{i}^{l}\circ
X_{j}^{l}+\left.\left(c_{jq}^{q}\hat
g^{ji}\right)\right|_{x_{0}}X_{i}^{l}+\left.\left[
\frac1{\sqrt{\hat\gamma}}X_{0}^{l}\left(\sqrt{\hat\gamma}\hat
g_{2}^{0i}\right)\right]\right|_{x_{0}}X_{i}^{l},
\end{equation}
where $X_{0}^{l}:=\partial/\partial r$.

The $G$-invariant measure $\mu_{2}$ on the space
$I'\times\left(G/K\right)$ has the form $\nu\otimes\mu$, where
$\nu=\phi(r)dr$ is the measure on the interval $I'$, and $\mu$ is
a $G$-invariant measure on the space $G/K$. The measure on the
space $I'\times G$, corresponding to $\mu_{2}$, has the form
$\tilde\mu_{2}=\nu\otimes\mu_{G}$, where $\mu_{G}$ is the
left-invariant measure on the group $G$, appropriately normalized.

Similarly to the section \ref{InvOperators}, we can define the
bijection $\zeta$ between the set of functions on the space
$I'\times(G/K)$ and the set of functions on the space $I'\times G$
that are invariant with respect to the right $K$-shifts. Denote by
$\mathcal{L}^{2}\left(I'\times G,K,\tilde\mu_{2}\right)$ the
Hilbert space of square integrable $K$-invariant functions on
$I'\times G$ with respect to the measure $\tilde\mu_{2}$ and the
right $K$-shifts. Thus we obtain the following isometry of Hilbert
spaces:
$$
\zeta: \mathcal{L}^{2}\left(M,\mu_{2}\right)\rightarrow
\mathcal{L}^{2}\left(I'\times G,K,\tilde\mu_{2}\right),
$$
and also
$\tilde\bigtriangleup_{g_{2}}\circ\zeta=\zeta\circ\bigtriangleup_{g_{2}}$.

\section{Two-point Hamiltonian for the general compact two-point
homogeneous space}\label{generaltwobody}
\markright{\ref{generaltwobody} Two point Hamiltonian for the
general case} In this section we shall obtain the concrete
expression for the two-point Hamiltonian of the form (\ref{LapUp})
on the general compact two-point homogeneous space. Let
\begin{equation}\label{fields}
L,X_{\lambda,i},Y_{\lambda,i},X_{2\lambda,j},Y_{2\lambda,j},\;i=1,\dots,q_{1},
j=1,\dots,q_{2}
\end{equation}
be the Killing vector fields on the space $Q$, corresponding to
the elements of the algebra $\mathfrak{g}$ from the proposition
\ref{prop3}
\begin{equation}\label{elements}
\Lambda,e_{\lambda,i},f_{\lambda,i},e_{2\lambda,j},f_{2\lambda,j},\;i=1,\dots,q_{1},
j=1,\dots,q_{2},
\end{equation}
with respect to the left action of the group $G$ on the space $Q$.
Define the curve $\hat\gamma$ on the space $Q$ by the formula
$\hat\gamma(s)=\exp\left(\dfrac{s}{R}\Lambda\right)x_{0}$. This
curve will be the necessary geodesic $\tilde\gamma$ from the
previous section according to the following proposition.
\begin{proposit}
\label{scal2}
\begin{enumerate}
\item Among all possible pairwise scalar products of fields (\ref{fields})
on the curve $\hat\gamma$ the nonzero products are the followings:
\begin{equation}\label{scal21}
\left.g(L,L)\right|_{\hat\gamma}=R^{2},
\end{equation}
\begin{equation}\label{scal22}
\left.g(X_{\lambda,i},X_{\lambda,i})\right|_{\hat\gamma}=\frac{R^{2}}2\left(1+\cos\frac{s}{R}
\right),\;i=1,\dots,q_{1},
\end{equation}
\begin{equation}\label{scal23}
\left.g(X_{\lambda,i},Y_{\lambda,i})\right|_{\hat\gamma}=-\frac{R^{2}}2\sin\frac{s}{R},
\;i=1,\dots,q_{1},
\end{equation}
\begin{equation}\label{scal24}
\left.g(Y_{\lambda,i},Y_{\lambda,i})\right|_{\hat\gamma}=\frac{R^{2}}2\left(1-\cos\frac{s}{R}
\right),\;i=1,\dots,q_{1},
\end{equation}
\begin{equation}\label{scal25}
\left.g(X_{2\lambda,i},X_{2\lambda,i})\right|_{\hat\gamma}=\frac{R^{2}}2\left(1+\cos\frac{2s}{R}
\right),\;i=1,\dots,q_{2},
\end{equation}
\begin{equation}\label{scal26}
\left.g(X_{2\lambda,i},Y_{2\lambda,i})\right|_{\hat\gamma}=-\frac{R^{2}}2\sin\frac{2s}{R},
\;i=1,\dots,q_{2},
\end{equation}
\begin{equation}\label{scal27}
\left.g(Y_{2\lambda,i},Y_{2\lambda,i})\right|_{\hat\gamma}=\frac{R^{2}}2\left(1-\cos\frac{2s}{R}
\right),\;i=1,\dots,q_{2};
\end{equation}
\item $\hat\gamma(s)=\tilde\gamma(s),\; s\in [0,\diam Q)$.
\end{enumerate}
\end{proposit}
\begin{proof}
 By construction, $\dfrac{L}{R}$ is the vector field tangent to the curve $\hat\gamma(s)$. Since
$$
\left.\frac{d}{ds}g(L,L)\right|_{\hat\gamma(s)}=\frac2Rg([L,L],L)=0,
$$
we have
$$
\left.g(\frac1RL,\frac1RL)\right|_{\hat\gamma(s)}\equiv
\left.g(\frac1RL,\frac1RL)\right|_{{\hat\gamma(0)}}=(\frac1R\Lambda,\frac1R\Lambda)=1
$$
which is equivalent to (\ref{scal21}), and so the parameter $s$ is
the natural parameter on the curve $\hat\gamma$. Using the
equality
$$
\left.\frac{d}{ds}g(X,Y)\right|_{\hat\gamma(s)}=\left.\frac{L}{R}\left(g(X,Y)\right)
\right|_{{\hat\gamma(s)}}=\left.\frac{1}{R}(g([L,X],Y)\right|_{{\hat\gamma(s)}}+
\left.\frac{1}{R}(g(X,[L,Y])\right|_{{\hat\gamma(s)}},
$$
where $X,Y$ are some vector fields on the curve $\hat\gamma$, the
relations (\ref{commSpesial1}) and the connection of the metric
$\left.g(\cdot,\cdot)\right|_{T_{x_{0}}Q}$ with the scalar product
$(\cdot,\cdot)$ on the algebra $\mathfrak{g}$, we obtain the
system of linear differential equations with initial conditions
with respect to all possible pairwise scalar products of the
fields (\ref{fields}) on the curve $\hat\gamma$. This system
decomposes to the set of easily solvable subsystems. For example,
one obtains
$$
\left.\frac{d}{ds}g(X_{\lambda,i},X_{\lambda,i})\right|_{\hat\gamma(s)}=\frac2R
\left.g\left([L,X_{\lambda,i}],X_{\lambda,i}\right)
\right|_{{\hat\gamma(s)}}=\frac1R
\left.g(Y_{\lambda,i},X_{\lambda,i})\right|_{{\hat\gamma(s)}},
$$
\begin{align*}
\left.\frac{d}{ds}g(Y_{\lambda,i},X_{\lambda,i})\right|_{\hat\gamma(s)}&=\frac1R
\left.g\left([L,Y_{\lambda,i}],X_{\lambda,i}\right)\right|_{{\hat\gamma(s)}}+\frac1R
\left.g\left(Y_{\lambda,i},[L,X_{\lambda,i}]\right)
\right|_{{\hat\gamma(s)}}\\&=-\frac1{2R}\left.g(X_{\lambda,i},X_{\lambda,i})
\right|_{{\hat\gamma(s)}}+\frac1{2R}\left.g(Y_{\lambda,i},Y_{\lambda,i})
\right|_{{\hat\gamma(s)}},
\end{align*}
$$
\left.\frac{d}{ds}g(Y_{\lambda,i},Y_{\lambda,i})\right|_{\hat\gamma(s)}=\frac2R
\left.g\left([L,Y_{\lambda,i}],X_{\lambda,i}\right)
\right|_{{\hat\gamma(s)}}=-\frac1R
\left.g(X_{\lambda,i},Y_{\lambda,i})\right|_{{\hat\gamma(s)}}.
$$
Taking into account the initial conditions given by
$$
\left.g(X_{\lambda,i},X_{\lambda,i})\right|_{\hat\gamma(0)}=(e_{\lambda,i},e_{\lambda,i})=R^{2},
\;\left.g(X_{\lambda,i},Y_{\lambda,i})\right|_{\hat\gamma(0)}=
\left.g(Y_{\lambda,i},Y_{\lambda,i})\right|_{\hat\gamma(0)}=0,\;i=1,\dots,q_{1},
$$
 (valid due to the formula
$\left.Y_{\lambda,i}\right|_{\hat\gamma(0)}=0$), we obtain
(\ref{scal22})-(\ref{scal24}). Other required formulae of the
first statement can be obtained in a similar way.

Let us prove the equality $\hat\gamma(s)=\tilde\gamma(s)$. It is
sufficient to show that
$\left.\nabla_{L}L\right|_{\hat\gamma(s)}=0$, since the parameters
of the curves $\hat\gamma(s),\tilde\gamma(s)$ are natural.
Formulae (\ref{conpart}), (\ref{commSpesial1}) and the first
statement of the proposition \ref{scal2} imply
\begin{align}\label{geod}
\left.g(\nabla_{L}L,X_{\lambda,i})\right|_{\hat\gamma(s)}&=
\left.g(L,[L,X_{\lambda,i}])\right|_{\hat\gamma(s)}=
\frac12\left.g(L,Y_{\lambda,i})\right|_{\hat\gamma(s)}=0,\;i=1,\dots,q_{1},\nonumber\\
\left.g(\nabla_{L}L,X_{2\lambda,j})\right|_{\hat\gamma(s)}&=
\left.g(L,[L,X_{2\lambda,j}])\right|_{\hat\gamma(s)}=
\left.g(L,Y_{2\lambda,j})\right|_{\hat\gamma(s)}=0,\;j=1,\dots,q_{2},\\
\left.g(\nabla_{L}L,L)\right|_{\hat\gamma(s)}&=\left.g(L,[L,L])\right|_{\hat\gamma(s)}=0.
\nonumber
\end{align}
Due to the first statement of this proposition the vector fields
$$L,X_{\lambda,i},X_{2\lambda,j},\;i=1,\dots,q_{1},j=1,\dots,q_{2}$$
form a moving frame in the tangent spaces $T_{\hat\gamma(s)}Q$ as
$s\in [0,\diam Q)$, since the matrix of their pairwise scalar
products in these spaces is nonsingular. Thus, due to (\ref{geod})
we have: $\left.\nabla_{L}L\right|_{\hat\gamma(s)}\equiv 0,\;s\in
[0,\diam Q]$.
\end{proof}
\begin{Rem}\label{geodesicsS}
It was mentioned above in  section \ref{twopoint} that the
decomposition $\mathfrak{g}=\mathfrak{p}\oplus\mathfrak{k}$ is
uniquely determined  by the point $x_{0}$. Therefore, due to
proposition \ref{scal2} and the isotropy of the space $Q$ all
nonzero elements of the space $\mathfrak{p}$ have the following
property: the trajectories of all one-parameter subgroups
corresponding to these elements and passing through the point
$x_{0}$ are geodesics. In particular it holds for the elements
$e_{\lambda,i},e_{2\lambda,j},\,i=1,\dots,q_{1},j=1,\dots,q_{2}$.
\end{Rem} Let us rename some notations to simplify the consideration of the
space $M=Q\times Q$. Now, let
\begin{align*}
L&=L^{(1)}+L^{(2)},\;X_{\lambda,i}=X_{\lambda,i}^{(1)}\oplus
X_{\lambda,i}^{(2)},\;Y_{\lambda,i}=Y_{\lambda,i}^{(1)}\oplus
Y_{\lambda,i}^{(2)},\,i=1,\dots,q_{1},\\
X_{2\lambda,j}&=X_{2\lambda,j}^{(1)}\oplus
X_{2\lambda,j}^{(2)},\;Y_{2\lambda,j}=Y_{2\lambda,j}^{(1)}\oplus
Y_{2\lambda,j}^{(2)},\,j=1,\dots,q_{2}
\end{align*}
be the decomposition of Killing vector fields on the space $M$,
which correspond to the elements
$\Lambda,e_{\lambda,i},f_{\lambda,i},e_{2\lambda,j},f_{2\lambda,j}$
and the decomposition $TM=TQ\oplus TQ$. Let $\gamma(p)$ be a curve
on the space $M$, constructed according to section \ref{hom} with
respect to the geodesic $\tilde\gamma$, and $X_{0}$ be the vector
field on the space $M$ constructed therein. Let $s_{1}(p)=\alpha
p,\;s_{2}(p)=-\beta p,\,\alpha,\beta\in (0,1),\,
\alpha+\beta=1,\;p=:2R\arctan r,\;r\in I'$, where $I'=(0,\infty)$
as $Q\ne\mathbf{P}^{n}(\mathbb{R})$ and $I'=(0,1)$ as
$Q=\mathbf{P}^{n}(\mathbb{R})$. Then
\begin{equation}\label{X0}
X_{0}=\frac d{dr}x_{p(r)}=\frac2{1+r^{2}}\left(\alpha
L^{(1)}-\beta L^{(2)}\right),
\end{equation}
since $\pi^{*}_{k}g(L,L)=R^{2},\;k=1,2$ and $\tilde\gamma(s)$ is
the normal parametrization of $\tilde\gamma$. Let us show that the
vector fields
\begin{equation}\label{basistwobody}
X_{0},\;L,\;X_{\lambda,i},\;Y_{\lambda,i},\;X_{2\lambda,j},\;Y_{2\lambda,j},\,i=1,\dots,q_{1},
j=1,\dots,q_{2}
\end{equation}
form a moving frame in a neighborhood of the curve
$\gamma(p),\;p\in (0,\diam Q)$. To prove this, we shall find the
matrix $\Gamma$ of pairwise scalar products of these fields on the
curve $\gamma$. Since $\pi^{*}_{k}g(L,L)=R^{2},\;k=1,2$, one has
\begin{align*}
\left.g_{2}(X_{0},X_{0})\right|_{\gamma}&=\left.g_{2}\left(\frac2{1+r^{2}}\left(\alpha
L^{(1)}-\beta L^{(2)}\right),\frac2{1+r^{2}}\left(\alpha
L^{(1)}-\beta
L^{(2)}\right)\right)\right|_{\gamma}\nonumber\\&=\frac{4R^{2}}{(1+r^{2})^{2}}(\alpha^{2}
m_{1}+\beta^{2}m_{2})=:a,\\ \left.g_{2}(L,X_{0})\right|_{\gamma}&=
\left.g_{2}\left(L^{(1)}+L^{(2)},\frac2{1+r^{2}}\left(\alpha
L^{(1)}-\beta L^{(2)}\right)\right)\right|_{\gamma}\nonumber\\
&=\frac{2R^{2}}{1+r^{2}}(\alpha m_{1}-\beta m_{2})=:b,\\
\left.g_{2}(L,L)\right|_{\gamma}&=(m_{1}+m_{2})R^{2}=:c.
\end{align*} Due to (\ref{X0}) and the orthogonality of the fields
$L^{(k)},\;k=1,2$ with respect to all fields
\begin{equation*}
X_{\lambda,i}^{(k)},\;Y_{\lambda,i}^{(k)},\;X_{2\lambda,j}^{(k)},\;Y_{2\lambda,j}^{(k)},\;
i=1,\dots,q_{1},j=1,\dots,q_{2},k=1,2
\end{equation*}
we obtain the orthogonality of the fields $X_{0},L$ with respect
to the fields
\begin{equation}
\label{fieldstwobody}
X_{\lambda,i},\;Y_{\lambda,i},\;X_{2\lambda,j},\;Y_{2\lambda,j},\;
i=1,\dots,q_{1},j=1,\dots,q_{2}.
\end{equation}
The proposition \ref{scal2} implies that among all possible
pairwise scalar products of the fields (\ref{fieldstwobody}) only
products $(X_{\lambda,i},Y_{\lambda,i}),\;i=1,\dots,q_{1}$ and
$(X_{2\lambda,j},Y_{2\lambda,j}),\;j=1,\dots,q_{2}$ can be
nonzero. By simple calculations, taking into account
(\ref{scal22})-(\ref{scal27}), we obtain
\begin{align*}
\left.g_{2}(X_{\lambda,i},X_{\lambda,i})\right|_{\gamma}&=R^{2}\left(m_{1}
\cos^{2}(\alpha\arctan r)+m_{2}\cos^{2}(\beta\arctan
r)\right)=:d,\\
\left.g_{2}(X_{\lambda,i},Y_{\lambda,i})\right|_{\gamma}&=R^{2}\left(-m_{1}
\sin(\alpha\arctan r)\cos(\alpha\arctan
r)\right.\nonumber\\&\left.+m_{2}\sin(\beta\arctan
r)\cos(\beta\arctan r)\right)=:h,\\
\left.g_{2}(Y_{\lambda,i},Y_{\lambda,i})\right|_{\gamma}&=R^{2}\left(m_{1}
\sin^{2}(\alpha\arctan r)+m_{2}\sin^{2}(\beta\arctan
r)\right)=:f,\;i=1,\dots,q_{1},\\
\left.g_{2}(X_{2\lambda,j},X_{2\lambda,j})\right|_{\gamma}&=R^{2}\left(m_{1}
\cos^{2}(2\alpha\arctan r)+m_{2}\cos^{2}(2\beta\arctan
r)\right)=:u,\\
\left.g_{2}(X_{2\lambda,j},Y_{2\lambda,j})\right|_{\gamma}&=R^{2}\left(-m_{1}
\sin(2\alpha\arctan r)\cos(2\alpha\arctan
r)\right.\nonumber\\&\left.+m_{2}\sin(2\beta\arctan
r)\cos(2\beta\arctan r)\right)=:w,\\
\left.g_{2}(Y_{2\lambda,j},Y_{2\lambda,j})\right|_{\gamma}&=R^{2}\left(m_{1}
\sin^{2}(2\alpha\arctan r)+m_{2}\sin^{2}(2\beta\arctan
r)\right)=:v,\;j=1,\dots,q_{2}.
\end{align*}
Thus we conclude that the matrix
$\Gamma=\left.g_{2}\right|_{\gamma}$ has a block structure with
the blocks:
$$
\begin{pmatrix} a & b \\ b & c\end{pmatrix}\,\mbox{ one time, }
\begin{pmatrix} d & h \\ h & f\end{pmatrix}\,- q_{1}\,\mbox{times and }
\begin{pmatrix} u & w \\ w & v\end{pmatrix}\,- q_{2}\,\mbox{times.}
$$
 We have therefore
$\det\Gamma=(ac-b^{2})(df-h^{2})^{q_{1}}(uv-w^{2})^{q_{2}}$. It is
easy to show that
$$
ac-b^{2}=\frac{4R^{4}m_{1}m_{2}}{(1+r^{2})^{2}},\,
df-h^{2}=\frac{R^{4}m_{1}m_{2}r^{2}}{1+r^{2}},\,
uv-w^{2}=\frac{4R^{4}m_{1}m_{2}r^{2}}{(1+r^{2})^{2}}.
$$
Thus
\begin{align}\label{detHgeneral}
\det\Gamma &
=\frac{4^{1+q_{2}}\left(R^4m_1m_2\right)^{1+q_{1}+q_{2}}r^{2(q_{1}+q_{2})}}
{(1+r^2)^{2+q_{1}+2q_{2}}},
\end{align}
\begin{align*}
\begin{pmatrix} a & b \\ b & c \end{pmatrix}^{-1}&=\frac1{4R^{2}m_{1}m_{2}}
\begin{pmatrix} (1+r^{2})^{2}(m_{1}+m_{2}) & 2(1+r^{2})(m_{1}\alpha-m_{2}\beta) \\
2(1+r^{2})(m_{1}\alpha-m_{2}\beta) &
4(m_{1}\alpha^{2}+m_{2}\beta^{2})
\end{pmatrix},\\
\begin{pmatrix} d & h \\ h & f\end{pmatrix}^{-1}&=
\begin{pmatrix} D_{s} & E_{s} \\ E_{s} & F_{s}\end{pmatrix},\,
\begin{pmatrix} u & w \\ w & v\end{pmatrix}^{-1}=
\begin{pmatrix} C_{s} & B_{s} \\ B_{s} & A_{s}\end{pmatrix},\,\mbox{where}
\end{align*}
\begin{align*}
D_{s}&=\frac{1+r^{2}}{m_{1}m_{2}R^{2}r^{2}}\left(m_1\sin^{2}(\alpha\arctan(r))+m_2\sin^{2}
(\beta\arctan(r))\right),\nonumber \\
F_{s}&=\frac{1+r^{2}}{m_{1}m_{2}R^{2}r^{2}}\left(m_1\cos^{2}(\alpha\arctan(r))+m_2\cos^{2}
(\beta\arctan(r))\right),\nonumber \\
E_{s}&=\frac{1+r^{2}}{2m_{1}m_{2}R^{2}r^{2}}\left(m_1\sin(2\alpha\arctan(r))-
m_2\sin(2\beta\arctan(r))\right),\nonumber \\
C_{s}&=\frac{(1+r^{2})^{2}}{4m_{1}m_{2}R^{2}r^{2}}\left(m_1\sin^{2}(2\alpha\arctan(r))+
m_2\sin^{2}(2\beta\arctan(r))\right),\\
A_{s}&=\frac{(1+r^{2})^{2}}{4m_{1}m_{2}R^{2}r^{2}}\left(m_1\cos^{2}(2\alpha\arctan(r))+
m_2\cos^{2}(2\beta\arctan(r))\right),\nonumber \\
B_{s}&=\frac{(1+r^{2})^{2}}{8m_{1}m_{2}R^{2}r^{2}}\left(m_1\sin(4\alpha\arctan(r))-
m_2\sin(4\beta\arctan(r))\right).\nonumber
\end{align*} In view of the formula (\ref{detHgeneral}), the fields
(\ref{basistwobody}) form a moving frame on the curve
$\gamma(p),\;p\in (0,\diam Q)$ and the condition \ref{B} is
satisfied.  Let
\begin{equation}
\label{repere}
L^{l},X_{\lambda,i}^{l},Y_{\lambda,i}^{l},X_{2\lambda,j}^{l},Y_{2\lambda,j}^{l},\,
i=1,\dots,q_{1},j=1,\dots,q_{2}
\end{equation}
 be left-invariant vector fields on the group $G$, corresponding to elements (\ref{elements})
of the algebra $\mathfrak{g}$ and $X_{0}^{l}=\partial/\partial r$
the vector field on $I'$. We consider the corresponding fields on
the space $I'\times G$ saving the notations. The field $X_{0}$
commutes with all fields (\ref{basistwobody}). So, due to the
proposition \ref{prop3}, the expansion of a commutator $[X,Y]$,
where X,Y are any elements of the frame (\ref{basistwobody}), by
the same frame, does not include $X,Y$. Thus, the second term in
the lift of the two-body Hamiltonian on the space $I'\times G$ in
accordance with (\ref{LapUp}) vanishes, since $c_{jq}^{q}=0$ (even
without summation over $q$). Consequently, this expression has the
form:
\begin{align}
\label{HupGeneral} \tilde H_0
&=-\frac{(1+r^2)^{1+\frac{q_{1}}2+q_{2}}}{8mR^2r^{q_{1}+q_{2}}}\pd1{r}\circ\left(
\frac{r^{q_{1}+q_{2}}}{(1+r^2)^{\frac{q_{1}}2+q_{2}-1}}\pd1{r}\right)-
\frac{(m_{1}\alpha-m_{2}\beta)(1+r^{2})^{1+\frac{q_{1}}2+q_{2}}}{4m_{1}m_{2}R^{2}
r^{q_{1}+q_{2}}}\notag \\ &\cdot
\left\{\pd1{r},\frac{r^{q_{1}+q_{2}}}{(1+r^{2})^{\frac{q_{1}}2+q_{2}}}
L^{l}\right\}-
\frac{m_{1}\alpha^{2}+m_{2}\beta^{2}}{2m_{1}m_{2}R^{2}}\left(L^{l}\right)^{2}-
\frac12\sum\limits_{i=1}^{q_{1}}\left(D_{s}\left(X_{\lambda,i}^{l}\right)^2
+F_{s}\left(Y_{\lambda,i}^{l}\right)^2\right.\notag \\
&+\left.E_{s}\left\{X_{\lambda,i}^{l},
Y_{\lambda,i}^{l}\right\}\right)-
\frac12\sum\limits_{j=1}^{q_{2}}\left(C_{s}\left(X_{2\lambda,j}^{l}\right)^2
+A_{s}\left(Y_{2\lambda,j}^{l}\right)^2+B_{s}\left\{X_{2\lambda,j}^{l},
Y_{2\lambda,j}^{l}\right\}\right),
\end{align}
where $\{X,Y\}=X\circ Y+Y\circ X$ is the anticommutator of $X$ and
$Y$, and $m:=\dfrac{m_{1}}{m_{2}}$.

According to section \ref{hom}, the lift of the measure, generated
by the metric $g_{2}$, on the space $I'\times G$ has the form
$\tilde\mu_{2}=\nu\otimes\mu_G$, where $\nu=\sqrt{\det\Gamma}dr$
is the measure on $I'$, and $\mu_G$ is the biinvariant measure on
the group $G$. Changing, if necessary, the normalization we get
$\nu=r^{q_{1}+q_{2}}dr/(1+r^{2})^{1+\frac{q_{1}}2+q_{2}}$. The
calculations above can be summarized in the following theorem.
\begin{theore}
The quantum two-body Hamiltonian on a compact two-point
homogeneous space $Q$ with the isometry group $G$ can be
considered as the differential operator $\tilde H_{0}+U(r)$ (where
the operator $\tilde H_{0}$ on the space $I'\times G$ is given by
the formula (\ref{HupGeneral})), where $I'=(0,1)$ in the case
$Q=\mathbf{P}^{n}(\mathbb{R})$ and $I'=(0,\infty)$ in other cases,
$\alpha,\beta\in (0,1),\,\alpha+\beta=1$. Its domain of definition
is dense in the space $\mathcal{L}^{2}\left(I'\times
G,K,\tilde\mu_{2}\right)$, consisting of all square integrable
$K$-invariant functions on $I'\times G$, with respect to the
measure $\tilde\mu_{2}$ and the right $K$-shifts.
\end{theore} \section{Two-point Hamiltonian for the general noncompact two-point
homogeneous space}\label{noncompact} \markright{\ref{noncompact}
Two point Hamiltonian on compact space.} Noncompact two-point
homogeneous spaces of types 7,8,9,10 are analogous to the compact
two-point homogeneous spaces of types 2,4,5,6, respectively. In
particular, it means that Lie algebras $\mathfrak{g}$ of symmetry
groups of analogous spaces are different real forms of a simple
complex Lie algebra. The transition from one such real form to
another can be done by multiplying the subspace $\mathfrak{p}$
from the decomposition
$\mathfrak{g}=\mathfrak{k}\oplus\mathfrak{p}$ by the imaginary
unit $\mathbf{i}$ (or by $-\mathbf{i}$). In the space $M=Q\times
Q$ this transition corresponds to the change $r\rightarrow
\mathbf{i}r,\,R\rightarrow \mathbf{i}R$ \cite{Shch98}. For
example, on  $\mathbf{ S}^{2}$ we have the transition from the
metric:
$$
\frac{4R^{2}(dr^{2}+r^{2}d\phi^{2})}{(1+r^{2})^{2}},\,r\in[0,\infty],\,\phi\in
\mathbb{R}\mod 2\pi
$$
to the metric:
$$
\frac{4R^{2}(dr^{2}-r^{2}d\phi^{2})}{(1-r^{2})^{2}},\,r\in[0,1),\,\phi\in
\mathbb{R}\mod 2\pi
$$
on the space $\mathbf{H}^{2}(\mathbb{R})$. It is clear that the
analogous spaces have equal multiplicities $q_{1}$ and $q_{2}$.
Thus, changing variables in (\ref{HupGeneral}) as
$$r\rightarrow \mathbf{i}r,\,R\rightarrow \mathbf{i}R,\,
X_{\lambda,i}^{l}\rightarrow
-\mathbf{i}X_{\lambda,i}^{l},\,X_{2\lambda,j}^{l}\rightarrow
-\mathbf{i}X_{2\lambda,j}^{l},$$ we obtain
\begin{theore}
The quantum two-body Hamiltonian on a noncompact two-point
homogeneous space $Q$ with the isometry group $G$ can be
considered as the differential operator
\begin{align}
\label{HupNoncompact} \tilde H
&=-\frac{(1-r^2)^{1+\frac{q_{1}}2+q_{2}}}{8mR^2r^{q_{1}+q_{2}}}\pd1{r}\circ\left(
\frac{r^{q_{1}+q_{2}}}{(1-r^2)^{\frac{q_{1}}2+q_{2}-1}}\pd1{r}\right)-
\frac{(m_{1}\alpha-m_{2}\beta)(1-r^{2})^{1+\frac{q_{1}}2+q_{2}}}{4m_{1}m_{2}R^{2}
r^{q_{1}+q_{2}}}\notag \\ &\cdot
\left\{\pd1{r},\frac{r^{q_{1}+q_{2}}}{(1-r^{2})^{\frac{q_{1}}2+q_{2}}}
L^{l}\right\}-
\frac{m_{1}\alpha^{2}+m_{2}\beta^{2}}{2m_{1}m_{2}R^{2}}\left(L^{l}\right)^{2}-
\frac12\sum\limits_{i=1}^{q_{1}}\left(D_{h}\left(X_{\lambda,i}^{l}\right)^2
+F_{h}\left(Y_{\lambda,i}^{l}\right)^2\right.\notag \\
&+\left.E_{h}\left\{X_{\lambda,i}^{l},
Y_{\lambda,i}^{l}\right\}\right)-
\frac12\sum\limits_{j=1}^{q_{2}}\left(C_{h}\left(X_{2\lambda,j}^{l}\right)^2
+A_{h}\left(Y_{2\lambda,j}^{l}\right)^2+B_{h}\left\{X_{2\lambda,j}^{l},
Y_{2\lambda,j}^{l}\right\}\right)+U(r),
\end{align}
where
\begin{align}
D_{h}&=\frac{1-r^{2}}{m_{1}m_{2}R^{2}r^{2}}\left(m_1\sinh^{2}(\alpha\arctanh(r))+m_2\sinh^{2}
(\beta\arctanh(r))\right),\nonumber \\
F_{h}&=\frac{1-r^{2}}{m_{1}m_{2}R^{2}r^{2}}\left(m_1\cosh^{2}(\alpha\arctanh(r))+m_2\cosh^{2}
(\beta\arctanh(r))\right),\nonumber \\
E_{h}&=\frac{1-r^{2}}{2m_{1}m_{2}R^{2}r^{2}}\left(m_1\sinh(2\alpha\arctanh(r))-
m_2\sinh(2\beta\arctanh(r))\right),\nonumber \\
C_{h}&=\frac{(1-r^{2})^{2}}{4m_{1}m_{2}R^{2}r^{2}}\left(m_1\sinh^{2}(2\alpha\arctanh(r))+
m_2\sinh^{2}(2\beta\arctanh(r))\right),\\
A_{h}&=\frac{(1-r^{2})^{2}}{4m_{1}m_{2}R^{2}r^{2}}\left(m_1\cosh^{2}(2\alpha\arctanh(r))+
m_2\cosh^{2}(2\beta\arctanh(r))\right),\nonumber \\
B_{h}&=\frac{(1-r^{2})^{2}}{8m_{1}m_{2}R^{2}r^{2}}\left(m_1\sinh(4\alpha\arctanh(r))-
m_2\sinh(4\beta\arctanh(r))\right),\nonumber
\end{align}
acting on the space $I'\times G$, where $I'=(0,1)$. Its domain of
definition is dense in the space $\mathcal{L}^{2}\left(I'\times
G,K,\tilde\mu_{2}\right)$, consisting of all square-integrable
$K$-invariant functions on $I'\times G$, with respect to the
measure $\tilde\mu_{2} =\nu\otimes\mu_G$ and the right $K$-shifts.
Now $\nu=r^{q_{1}+q_{2}}dr/(1-r^{2})^{1+\frac{q_{1}}2+q_{2}}$ is
the measure on $I'$ and $\mu_G$ is biinvariant measure on $G$,
since $G$ is unimodular.
\end{theore}
The following remark is analogous to  remarks \ref{geodesicsS} and
\ref{GeodesicSubmanifolds}.
\begin{Rem}\label{geodesicsH}
The space $\mathfrak{a}\oplus\mathfrak{p}_{2\lambda}$ generates in
the space $Q$ a completely geodesic submanifold of the constant
sectional curvature $-R^{-2}$, isomorphic to the space ${\bf
H}^{q_{2}+1}(\mathbb{R})$.

If $q_{1}\neq 0$, the element $\Lambda$ and an arbitrary nonzero
element from the space $\mathfrak{p}_{\lambda}$ generate in $Q$ a
completely geodesic two dimensional submanifolds of constant
curvature $-R^{-2}$.

The trajectories of all one-parameter subgroups corresponding to
the elements of the space $\mathfrak{p}$, passing through the
point $x_{0}$, are geodesics. In particular, it holds for the
elements
$e_{\lambda,i},e_{2\lambda,j},\,i=1,\dots,q_{1},j=1,\dots,q_{2}$.
\end{Rem}

\section{The Hamiltonian function for the classical two-body
problem on two-point homogeneous spaces} \label{ClassicalSystems}
\markright{\ref{ClassicalSystems} The Hamiltonian function for the
classical two body problem}

We can derive the Hamiltonian functions of classical two-body
problems on two-point homogeneous spaces from (\ref{HupGeneral})
and (\ref{HupNoncompact}). These functions are defined on the
cotangent bundles $T^{*}(Q\times Q\backslash\diag)$ and are
polynomials of the second order on each fiber.

The $G$-action on the space $(Q\times Q)\backslash\diag$ can be
naturally lifted to the Poisson action on the space
$T^{*}((Q\times Q)\backslash\diag)$ \cite{Vinberg}, \cite{Arn2}.
It means that for any $X\in\mathfrak{g}$ there is a function
$p_{X}$ on the space $T^{*}((Q\times Q)\backslash\diag)\cong
T^{*}I'\times T^{*}(G/K)$ linear on fibers. The Hamiltonian vector
field, corresponding to this function, coincides with the lift of
the Killing vector field for $X$ onto the cotangent bundle. All
such functions are integrals for all $G$-invariant Hamiltonian
systems on $T^{*}I'\times T^{*}(G/K)$ and can be considered as the
generalized momenta. The set of such functions is a Lie algebra
with respect to the Poisson bracket, and the correspondence
$X\rightarrow p_{X}$ is the isomorphism of Lie algebras. To obtain
the classical Hamiltonian functions from quantum Hamiltonians we
should change the left invariant vector fields and the operator
$\partial/\partial r$ to corresponding momenta, multiplied by the
imaginary unit in formulae (\ref{HupGeneral}) or
(\ref{HupNoncompact}). Denote the momentum, corresponding to the
operator $\partial/\partial r$, by $p_{r}$, and momenta
corresponding to the fields (\ref{repere}), by
\begin{equation}\label{impulses}
p_{L},p_{x,\lambda,i},p_{y,\lambda,i},p_{x,2\lambda,j},p_{y,2\lambda,j},\,
i=1,\dots,q_{1},j=1,\dots,q_{2}.
\end{equation} Then the Hamiltonian function for the classical two-body problem
on two-point compact homogeneous spaces has the form:
\begin{align}
\label{HupComClas} H_{s} &=\frac{(1+r^2)^{2}}{8mR^2}p_{r}^{2}+
\frac{(m_{1}\alpha-m_{2}\beta)(1+r^{2})}{4m_{1}m_{2}R^{2}}p_{r}p_{L}+
\frac{m_{1}\alpha^{2}+m_{2}\beta^{2}}{2m_{1}m_{2}R^{2}}p_{L}^{2}\notag\\
&+\frac12\sum\limits_{i=1}^{q_{1}}\left(D_{s}\left(p_{x,\lambda,i}\right)^2+F_{s}\left(p_{y,\lambda,i}\right)^2+2E_{s}
p_{x,\lambda,i}p_{y,\lambda,i}\right)\notag\\ &+
\frac12\sum\limits_{j=1}^{q_{2}}\left(C_{s}\left(p_{x,2\lambda,j}\right)^2
+A_{s}\left(p_{y,2\lambda,j}\right)^2+2B_{s}p_{x,2\lambda,j}p_{y,2\lambda,j}\right)+U(r),
\end{align}
and on the two-point noncompact homogeneous spaces other than the
Euclidean one, it has the form:
\begin{align}
\label{HupNoncomClas} H_{h} &=\frac{(1-r^2)^{2}}{8mR^2}p_{r}^{2}+
\frac{(m_{1}\alpha-m_{2}\beta)(1-r^{2})}{4m_{1}m_{2}R^{2}}p_{r}p_{L}+
\frac{m_{1}\alpha^{2}+m_{2}\beta^{2}}{2m_{1}m_{2}R^{2}}p_{L}^{2}\notag\\
&+\frac12\sum\limits_{i=1}^{q_{1}}\left(D_{h}\left(p_{x,\lambda,i}\right)^2+F_{h}\left(p_{y,\lambda,i}\right)^2+2E_{h}
p_{x,\lambda,i}p_{y,\lambda,i}\right)\notag\\ &+
\frac12\sum\limits_{j=1}^{q_{2}}\left(C_{h}\left(p_{x,2\lambda,j}\right)^2
+A_{h}\left(p_{y,2\lambda,j}\right)^2+2B_{h}p_{x,2\lambda,j}p_{y,2\lambda,j}\right)+U(r).
\end{align} This form of the Hamiltonian function is convenient for the
Marsden-Weinstein reduction. It is clear that this reduction acts
only on the second factor in the expansion of the phase space
$T^{*}I'\times T^{*}(G/K)$. The description of reduced spaces for
the space $T^{*}(G/K)$ with respect to this reduction was obtained
in \cite{Shchep1} in terms of the $\Ad^{*}_{G}$-orbits. Take an
arbitrary $\Ad^{*}_{G}$-orbit $\mathcal{O}$ and find its
submanifold $\mathcal{O}'$ annulled by the subalgebra
$\mathfrak{k}$. The quotient space $\tilde\mathcal{O}$ of
$\mathcal{O}'$ with respect to $\Ad^{*}_{K}$ action is isomorphic
to the reduced phase space for the space $T^{*}(G/K)$. Hence
reducing the Hamiltonian two-body system on two-point homogeneous
spaces we obtain the Hamiltonian system on the space
$T^{*}I'\times \tilde\mathcal{O}$.

Practically it means the following. Generalized momenta
(\ref{impulses}) corresponding to the elements of the basis in the
space $\mathfrak{p}\subset\mathfrak{g}$ can be considered as
linear functions on the annulator of the subalgebra $\mathfrak{k}$
in the space $\mathfrak{g}^{*}$ in view of the expansion
$\mathfrak{g}=\mathfrak{p}\oplus\mathfrak{k}$ and the isomorphism
$(\mathfrak{g}^{*})^{*}\cong\mathfrak{g}$. Therefore the momenta
(\ref{impulses}) themselves can be considered as functions on the
space $\tilde\mathcal{O}$. Combinations of these functions,
independent on the space $\tilde\mathcal{O}$, are coordinates on
$\tilde\mathcal{O}$ and their commutative relations define the
symplectic structure on the space $\tilde\mathcal{O}$.

\section{Mass center for two particles on two-point homogeneous
spaces}\label{centermass} \markright{\ref{centermass} Mass center
for two particles on two point homogeneous spaces}

The importance of the mass center concept for isolated system of
particles or a rigid body in Euclidean space stems from the
following properties:
\begin{enumerate}
\item it moves with a constant speed along a (geodesic) line for a classical mechanical
system;
\item variables corresponding to the mass center are separated from other variables both in
classical and quantum mechanical problems.
\end{enumerate}
These properties imply, in particular, that the (generally
complicated) motion of a system can be decomposed into the motion
of one point representing the center of mass, and the motion of
the system with respect to that point, often greatly simplifying
the problem. Under the action of external forces the center of
mass moves as if all forces act on the particle located at the
center of mass and having the mass equal to the total mass of the
system. An attempt to generalize the concept of the center of mass
to the curved two-point homogeneous Riemannian spaces encounters
difficulties related to the absence of nice dynamical properties
such as 1 and 2 above. It is natural to define the mass center for
the two particles on a two-point homogeneous Riemannian space as
the point on the geodesic interval connecting these particles that
divides the interval in definite ratio. If this ratio is equal to
the ratio of particle masses, we denote the corresponding mass
center by $R_{1}$.

However, even for spaces of constant sectional curvature, such a
mass center does not have property 1 \cite{Shch98}. For example,
consider two free particles on the sphere $\mathbf{S}^{2}$. Choose
two antipodal points on the sphere (poles), and the equator
connecting them. Let one point rest at the pole and another move
with the constant speed along the equator. Then any point on the
interval connecting those particles does not move along geodesic
unless this point coincides with one of the particles. The latter
is obviously senseless. Therefore for the mass center on a
two-point homogeneous Riemannian space we must rely on properties
different from the property 1.

\subsection{Existing mass center
concepts for spaces of a constant curvature}

The axiomatic approach to the concept of mass center was developed
in \cite{Galperin1},\cite{Galperin2}. Let
$\mathfrak{A}=\{(A_{i},m_{i})\}$ be a system (possibly empty) of
material points $A_{i}$ with masses $m_{i}$ in the space $Q$ of
constant sectional curvature, which corresponds to the types 2 or
9 according to the classification given in section \ref{twopoint}.
Denote by $\mathcal{A}$ the set of all such systems and by
$\mathcal{A}_{0}$ the subset of one-particle systems. For any
positive real number $\chi$ define the operation
$\chi\cdot\mathfrak{A}=\{(A_{i},\chi m_{i})\}$.
\begin{theore}[\cite{Galperin2}]\label{Galperin}
There is a unique map $\mathbb{U}$ of the set $\mathcal{A}$ onto
the set $\mathcal{A}_{0}$, satisfying the following axioms: 1)
$\mathbb{U}\left(\{(A_{1},m_{1})\}\right)=\{(A_{1},m_{1})\}$, 2)
$\mathbb{U}\left(\mathfrak{A}\cup\mathfrak{B}\right)
=\mathbb{U}\left(\mathbb{U}(\mathfrak{A})\cup\mathbb{U}(\mathfrak{B})\right)$;
3)
$\mathbb{U}(\chi\cdot\mathfrak{A})=\chi\cdot\mathbb{U}(\mathfrak{A})$;
4) $\mathbb{U}\circ q=q\circ\mathbb{U}$; 5) the map $\mathbb{U}$
is continuous with respect to the natural topology on the space
$\mathcal{A}$. Two systems are close to each other in this
topology, if their material points are pairwise close and have
similar masses. Points with small masses are close to the empty
set.

For the sphere $\mathbf{S}^{n}$ with unit curvature this map
$\mathbb{U}$ takes the system $\{(A_{1},m_{1}),(A_{2},m_{2})\}$ to
the material point (mass center), located on the geodesic interval
connecting the points $A_{1},A_{2}$ and dividing it in the ratio
$\dfrac{\rho_{1}}{\rho_{2}}$, as measured from the point $A_{1}$.
Besides,
$m_{1}\sin(\rho_{1})=m_{2}\sin(\rho_{2}),\,\rho_{1}+\rho_{2}=\rho$,
where $\rho$ is the distance between particles and
$\rho_{i},\,i=1,2$ is the distance between $i$-th particle and the
mass center. The mass of the mass center is assumed to be
$\cos(\rho_{1})m_{1}+\cos(\rho_{2})m_{2}$.

For the Lobachevski space $\mathbf{H}^{n}(\mathbb{R})$ with unit
curvature the map $\mathbb{U}$ is obtained by using the hyperbolic
functions $\sinh,\,\cosh$ instead of the corresponding
trigonometric functions $\sin,\,\cos$.
\end{theore} This approach to the definition of the center of mass
 corresponds to
the mass center concept in  flat space-time of special relativity
(SR) \cite{Galperin2}. In fact, for a given inertial frame of
reference, there exists a one to one correspondence between
possible particle velocities in SR and material points in the
space $\mathbf{H}^{3}(\mathbb{R})$, with masses equal to the rest
masses in SR. Therefore, a system $\mathfrak{A}\in\mathcal{A}$
corresponds to a system $\varsigma(\mathfrak{A})$ of moving
particles in SR. The total mass and momentum of the latter system
uniquely determine the rest mass and velocity of some effective
particle $\Xi$ in SR. This particle determines the mass center
$\varsigma^{-1}(\Xi)$ of the system $\mathfrak{A}$ in the space
$\mathbf{H}^{3}(\mathbb{R})$. We denote the mass center defined in
this way by $R_{2}$.

It is clear that this definition of a mass center can be easily
generalized to systems with a distributed mass.

Note that the mass center $R_{2}$ of two particles with equal
masses located at the diametrically opposite points of a sphere
has an arbitrary position on the equator and the null mass, which
is equivalent to the empty set.

The definition of the mass center $R_{2}$ seems to be quite
natural. Unfortunately, no ``good'' dynamical properties are known
for it. In order to find the mass center with such properties, we
can try to search for a pure geometrical mass center without any
mass. In this case we need not be concerned about the validity of
axioms 2 and 4 of the theorem \ref{Galperin}, and thus have more
freedom. This approach to the mass center concept concerning the
free motion on spaces
$\mathbf{S}^{n},\,\mathbf{H}^{n}(\mathbb{R}),\,n=2,3$ was
developed to various degrees of generality in
\cite{Zitt}-\cite{Salvai}. Consider the following definition of a
mass center. Let $Q=\mathbf{H}^{n}(\mathbb{R}),\,n=2,3$. Define a
rigid body in $Q$ by a nonnegative density function
$\varrho(x),\,x\in Q$ with a compact connected support, and
consider the function
\begin{equation}\label{FunMassS}
\Upsilon(x)=\int_{Q}\sinh^{2}(\rho(x,y))\varrho(y)d\mu,
\end{equation}
where $\mu$ is the measure on the space $Q$, generated by the
Riemannian metric. This function has a unique minimum and the
coordinate of this minimum can be chosen as a definition of the
center of mass $R_{3}$ for the rigid body. It is clear that the
similar definition is also suitable for a system of particles.

Unlike the center of mass $R_{2}$, the mass center $R_{3}$ for two
particles is determined from the equations
$m_{1}\sinh(2\rho_{1})=m_{2}\sinh(2\rho_{2}),\,\rho_{1}+\rho_{2}=\rho$.
Here as before $\rho$ is the distance between the particles, and
$\rho_{i},\,i=1,2$ is the distance between the $i$-th particle and
the mass center located on the geodesic interval connecting the
particles.

There are three types of actions of one parameter subgroups
$\exp(tX),\, X\in\mathfrak{g},t\in\mathbb{R}$ of the group $G$ in
the hyperbolic space $Q$ \cite{BalVor}. The one parameter
subgroup, isomorphic to $\mathbb{S}^{1}$, conserves all points of
a completely geodesic submanifold of codimension two and is called
{\it rotation} around some geodesic ({\it an axis of a rotation})
for $Q=\mathbf{H}^{3}(\mathbb{R})$ or around some point ({\it a
center of a rotation}) for $Q=\mathbf{H}^{2}(\mathbb{R})$. The
corresponding element $X$ is called {\it elliptic}. If a
one-parameter subgroup, isomorphic to $\mathbb{R}$, conserves some
geodesic then it is called a {\it transvection} along this
geodesic ({\it an axis of a transvection}). The corresponding
element $X$ is called {\it hyperbolic}. The last type of action of
a one-parameter subgroup is a parabolic action of $\mathbb{R}$. It
shifts points of $Q$ along the system of horocycles that are lines
orthogonal at each point to all geodesics having a common point on
the absolute. The corresponding element $X$ is called {\it
parabolic}.

Call a free movement of a rigid body a {\it free rotation} if all
points of this body move along trajectories of some rotation. Call
a free movement of a rigid body a {\it free transvection} if all
points of this body move along trajectories of some transvection.
The mass center $R_{3}$ has the following dynamical properties:
\begin{enumerate}
\item The free rotation of a rigid body around its mass center is possible in the space $\mathbf{H}^{n}(\mathbb{R})$.
If $n=2$, there is only one such rotation \cite{Nagy} and if $n=3$
there are three different rotations \cite{Salvai} around three
pairwise perpendicular axes passing through the mass center
$R_{3}$.
\item All possible transvections of a rigid body have axes passing through the mass center
$R_{3}$. For $n=2$ there are two such geodesics. For $n=3$ there
are three such geodesics, and they coincide with the axes of free
rotations.
\item The mass center $R_{3}$ is uniquely determined
 by any of the properties 1 or 2.
\item The velocities of all possible free rotations and transvections are constant.
\item There are no free movements of a rigid body along
horocycles \cite{Nagy}.
\end{enumerate} The situation for the spaces
$Q=\mathbf{S}^{n}(\mathbb{R}),\,n=2,3$ is analogous if we restrict
ourselves to  rigid bodies of  ``moderate'' sizes, i.e. if the
diameter of a rigid body is no more than $\pi R/4$ \cite{Zitt}.
This condition is required in order to differ transvections and
rotations of rigid bodies by the location of immovable points of
one parameter isometry subgroups with respect to the rigid body
itself, since all such subgroups of the isometry group ${\bf
SO}(n+1)$ are conjugated, and their trajectories in the space $Q$
are equivalent.

Note that most free movements of a rigid body in spaces of
constant sectional curvature do not correspond to the center of
mass $R_{3}$ movement along a geodesic even when this rigid body
is a homogeneous ball \cite{Zitt}.

\subsection{The connection of existing mass center concepts to the two-point Hamiltonian}

Consider now the connection of formulae (\ref{HupGeneral}),
(\ref{HupNoncompact}) obtained for the two-body Hamiltonian to the
mass center concepts. If we fix the parameter $\alpha$, then the
particle positions uniquely determine the location of the point
$\tilde\gamma(0)$ in the space $Q$ at every moment of time. This
point divides the geodesic interval $\tilde\gamma(t)$, $t\in
[s_{1}(s),s_{2}(s)]$ of
 length $s$ in the ratio $\dfrac{\alpha}{1-\alpha}$. The left-invariant
vector fields
\begin{equation}
L^{l},X_{\lambda,i}^{l},X_{2\lambda,j}^{l},\,i=1,q_{1},j=1,q_{2}
\end{equation}
on the group $G$ in formulae (\ref{HupGeneral}),
(\ref{HupNoncompact}) correspond to the basis of the space
$\mathfrak{p}\in\mathfrak{g}$. According to remarks
\ref{geodesicsS} and \ref{geodesicsH}, the trajectories of
one-parameter subgroups generated by those fields and passing
through the point $\tilde\gamma(0)$ are geodesics. The dynamical
approaches to the definition of a mass center considered above are
based on a possible movement of a mass center along geodesics of
the space $Q$. Therefore, dynamical properties of a point
representing a potential candidate for the mass center role can be
studied by identifying it with the point $\tilde\gamma(0)$. Such
an identification can always be achieved by choosing the parameter
$\alpha$ appropriately.

\begin{Def}\label{defcenter}
Let $Q_{2}\subset Q\times Q$ be a set of two particle positions
that correspond to the only one shortest path connecting
particles. A map from $Q_{2}$ to $Q$ is called the dynamical mass
center if it maps a two particle position from $Q_{2}$ to the
point on the geodesic interval connecting the particles that
divides the length of this interval in some ratio depending only
on particle masses. Besides, for any geodesic on $Q$ and any
interactive potential there should be some initial positions and
velocities of particles such that this point moves along this
geodesic with a constant speed. For brevity, we call the value of
this map the ``dynamical mass center''.
\end{Def} Note that this definition is appropriate for any Riemannian space
$Q$. For two-point homogeneous spaces the set $Q_{2}$ is open and
dense in $Q\times Q$. According to what was stated above and in
section \ref{ClassicalSystems}, the point $\tilde\gamma(0)$ moves
along a geodesic with a constant speed if and only if the
following equality holds:
\begin{equation}\label{proportional}
\pi_{2}(dH_{s,h})=\omega
dp_{L}+\sum\limits_{i=1}^{q_{1}}\omega'_{i}dp_{x,\lambda,i}+\sum\limits_{j=1}^{q_{2}}\omega''_{j}dp_{x,\lambda,j},
\end{equation}
where  $\pi_{2}(dH_{s,h})$ is the projection of the differential
$dH_{s,h}$ of the function (\ref{HupComClas}) or
(\ref{HupNoncomClas}) onto the tangent space to the second factor
of the expansion $T^{*}I'\times T^{*}(G/K)$,
$\omega,\omega'_{i},\omega''_{j}$ are some constants and
$\omega^{2}+\sum_{i=1}^{q_{1}}(\omega'_{i})^{2}+\sum_{j=1}^{q_{2}}(\omega''_{j})^{2}\neq
0 $. It is clear that
\begin{align}
\pi_{2}(dH_{s,h})=\left(\frac{(m_{1}\alpha-m_{2}\beta)(1\pm
r^{2})}{4m_{1}m_{2}R^{2}}p_{r}+
\frac{m_{1}\alpha^{2}+m_{2}\beta^{2}}{m_{1}m_{2}R^{2}}p_{L}\right)dp_{L}\notag\\
+\sum\limits_{i=1}^{q_{1}}\left[\left(D_{s,h}p_{x,\lambda,i}+E_{s,h}p_{y,\lambda,i}\right)dp_{x,\lambda,i}+
\left(F_{s,h}p_{y,\lambda,i}+E_{s,h}p_{x,\lambda,i}\right)dp_{y,\lambda,i}\right]
\notag\\
+\sum\limits_{j=1}^{q_{2}}\left[\left(C_{s,h}p_{x,2\lambda,i}+B_{s,h}p_{y,2\lambda,i}\right)dp_{x,2\lambda,i}+
\left(A_{s,h}p_{y,2\lambda,i}+B_{s,h}p_{x,2\lambda,i}\right)dp_{y,2\lambda,i}\right].
\end{align}
In the case of an arbitrary potential $U(r)$ the variables $r$ and
$p_{r}$ (and also the functions
$A_{s,h},B_{s,h},C_{s,h},D_{s,h},E_{s,h},F_{s,h}$) can take
arbitrary values on a trajectory. Therefore the equality
(\ref{proportional}) is possible only if
$m_{1}\alpha-m_{2}\beta=0$ and
\begin{align}\label{impulsNull}
p_{L}=\const\neq 0,p_{x,\lambda,i}=p_{y,\lambda,i}=0,
p_{x,2\lambda,j}=p_{y,2\lambda,j}=0,\,i=1,\dots,q_{1},j=1,\dots,q_{2}.
\end{align}
In view of commutative relations (\ref{commSpesial1}) the
equalities (\ref{impulsNull}) are conserved on a trajectory of the
dynamical system. In this case $dH_{s,h}\sim dp_{L}$ and the
motion of both particles is along the common geodesic.

The equality $m_{1}\alpha-m_{2}\beta=0$ gives the ratio of the
distances $s_{1}$ and $s_{2}$:
$\dfrac{s_{1}}{s_{2}}=\dfrac{m_{2}}{m_{1}}$, which corresponds to
the mass center $R_{1}$. Thus only the mass center $R_{1}$
satisfies the definition \ref{defcenter}. Note the connection of
the mass center $R_{3}$ with the zeroes of coefficients $B_{s,h}$
and $E_{s,h}$. If $B_{s}=0$, we have
$m_{1}\sin(\dfrac{2s_{1}}R)=m_{2}\sin(\dfrac{2s_{2}}R)$ which
means the coincidence of the points $\tilde\gamma(0)$ and $R_{3}$.
According to remarks \ref{GeodesicSubmanifolds} and
\ref{geodesicsH}, the momenta $p_{x,2\lambda,j}$ and
$p_{y,2\lambda,j}$ for some fixed $j$ correspond to the
instantaneous motion of particles along a two-dimensional
completely geodesic submanifold of the constant curvature $\pm
R^{-2}$.

If $E_{s}=0$, we have
$m_{1}\sin(\dfrac{s_{1}}R)=m_{2}\sin(\dfrac{s_{2}}R)$. Due to
remarks \ref{GeodesicSubmanifolds} and \ref{geodesicsH}, momenta
$p_{x,\lambda,i}$ and $p_{y,\lambda,i}$ for fixed $i$ correspond
to the instantaneous motion of particles along a two dimensional
completely geodesic submanifold of the constant curvature $\pm
(2R)^{-2}$. Therefore, in this case also the point
$\tilde\gamma(0)$ corresponds to the mass center $R_{3}$.

Let us note in conclusion that by appropriately choosing  the
parameter $\alpha$, expressions (\ref{HupGeneral}) and
(\ref{HupNoncompact}) can be simplified such that coefficients
$m_{1}\alpha-m_{2}\beta,\,B_{s,h}$ or $E_{s,h}$  vanish. These
values of the parameter $\alpha$ correspond to the mass center
concepts $R_{1}$ and $R_{3}$. In Euclidean case the choice
$\alpha=\dfrac{m_{2}}{m_{1}+m_{2}}$ leads to the separation of the
variable $r$ from other variables in two-point Hamiltonian. On
two-point homogeneous spaces it is impossible to separate the
variable $r$ from other variables by means of the choice the
parameter $\alpha$.

\end{document}